\documentclass[runningheads,envcountsect]{llncs}

\usepackage[paperheight=235mm, paperwidth=155mm,textwidth=12.2cm,textheight=19.3cm,hmarginratio=1:1]{geometry}

\usepackage{amssymb}
\usepackage{bbding}

\usepackage[%
  bookmarks,
  unicode,
  colorlinks=true,
  allcolors=blue!65!black!90,
  breaklinks=true]{hyperref}

\usepackage{amsmath}

\usepackage{cleveref}
\Crefname{figure}{Fig.}{Figs.}
\crefname{figure}{fig.}{figs.}
\Crefname{tabular}{Tab.}{Tabs.}
\crefname{tabular}{tab.}{tabs.}
\Crefname{section}{Sec.}{Sects.}
\crefname{section}{sec.}{sects.}
\Crefname{appendix}{App.}{Apps.}
\crefname{appendix}{app.}{apps.}
\Crefname{equation}{Eq.}{Eqs.}
\crefname{equation}{eq.}{eqs.}
\creflabelformat{equation}{#2#1#3}
\Crefname{example}{Ex.}{Exs.}
\crefname{example}{ex.}{exs.}
\Crefname{definition}{Def.}{Defs.}
\crefname{definition}{def.}{defs.}
\Crefname{theorem}{Thm.}{Thms.}
\crefname{theorem}{thm.}{thms.}
\Crefname{convention}{Conv.}{Convs.}
\crefname{convention}{conv.}{convs.}

\usepackage{listings}

\allowdisplaybreaks

\newcommand{\sett}{\mathrm{set}} %
\newcommand{\M}{\mathcal{M}} %
\newcommand{\U}{\mathbb{U}} %
\newcommand{\F}{\mathcal{F}} %
\newcommand{\N}{\mathbb{N}} %
\newcommand{\R}{\mathbb{R}} %
\newcommand{\Ng}{\mathbb{N}^+} %
\renewcommand{\P}{\mathbf{P}} %
\newcommand{\E}{\mathbb{E}} %
\newcommand{\inp}{\mathrm{inp}}
\newcommand{\red}{\mathrm{red}}
\newcommand{\abstr}{\mathrm{abstr}}

\usepackage{xspace}
\newcommand{\eg}{e.g.\ }
\newcommand{\ie}{i.e.\ }
\newcommand{\etal}{et al.\xspace}
\newcommand{\wrt}{w.r.t.\xspace}

\newcommand{\set}[1]{\ensuremath{\{\,#1\,\}}}

\usepackage{physics}

\makeatletter
\def\orcidID#1{\textsuperscript{\,\smash{\protect\raisebox{-1.25pt}{\href{http://orcid.org/#1}{\protect\includegraphics[scale=.8]{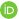}}}}}}
\makeatother

\newcommand{\isajot}{1pt} %
\newcommand{\isaindent}{0pt} %
\makeatletter
\newcommand{\isa@setup}{\setlength{\jot}{\isajot}\@fleqntrue\setlength{\@mathmargin}{\isaindent}}
\newenvironment{isaalign}{\isa@setup\align}{\endalign}
\newenvironment{isaalign*}{\isa@setup\csname align*\endcsname}{\csname endalign*\endcsname}
\makeatother

\usepackage{tikz}
\usepackage{tikz-cd}
\tikzset{squared node/.style={rectangle,draw, inner sep=0.4, minimum size=1.5em},
nodestyle/.style={squared node,text height=.8em,text depth=.25em}}

\usepackage{pgfplots}
\pgfplotsset{compat=1.18} %

\usepackage{xcolor}
\usepackage{graphicx}
\usepackage[most]{tcolorbox}
\definecolor{listingbgcolor}{HTML}{d9d9db}
\definecolor{verylightgray}{HTML}{f0f0f0}
\newcommand{\sverb}[1]{{\small \verb|#1|}}

\spnewtheorem{convention}{Convention}{\bfseries}{\upshape}
\renewenvironment{proof}[1][\proofname]{\par\trivlist\item[\hskip\labelsep{\itshape #1.}]\rmfamily\ignorespaces}{\qed\endtrivlist}

\begin{document}

\title{Path Abstraction for Markov Reward Models: An Isabelle Formalisation in a Free Monoid View}
\titlerunning{Path Abstraction for MRMs: A Free-Monoid Isabelle Formalisation}

\author{%
Arnd Hartmanns\orcidID{0000-0003-3268-8674}
\and
Robert Modderman$^{\text{\,(\raisebox{-1.6pt}{\Envelope})}}$\orcidID{0009-0002-9198-3809}
}
\authorrunning{A.\ Hartmanns and R.\ Modderman}
\institute{%
University of Twente, Enschede, The Netherlands
$\cdot$ \email{r.modderman@utwente.nl}
}

\maketitle

\begin{abstract}
The classical definition of discrete-time Markov chains maps pairs of states to transition probabilities.
An alternative, proposed in the context of path abstraction in probabilistic counterexample refinement, is to view them as a transition kernel on the \emph{free monoid} on the state space, mapping paths directly to path probabilities.
In this paper, we first extend the free monoid view and path abstraction algorithm to cumulative rewards on Markov reward models.
We then use Isabelle/HOL to formalise Markov chains and reward models in this view, exploiting the discrete finite-prefix semantics to work with Isabelle's standard HOL-Analysis library.
We mechanise the path abstraction algorithm and its key correctness properties of well-definedness and monotonic absorption, establishing the preservation of reachability probabilities and expected rewards.
Proving the convergence of these quantities on relevant path sets is aided by our use of the lightweight free monoid semantics in place of the standard measure-theoretic one used in established Isabelle formalisations of Markov models.
We obtain a verified abstract implementation of path abstraction in Isabelle/HOL, providing a foundation for refinement to efficient executable code and integration with probabilistic model checkers.
\end{abstract}

\section{Introduction}
\label{sec:introduction}

Markov chains and reward models are \emph{the} basic formalisms to model probabilistic reactive systems, widely used in dependability and performance evaluation~\cite{BHHK10}.
They also provide the semantics for various probabilistic programming languages~\cite{GHNR14}.
Efficient tools use probabilistic and statistical model checking~\cite{AP18,BAFK18,BHK+20} to check properties of interest, \eg to determine whether $\phi^G_{\leq p}$---a PCTL~\cite{HJ94} formula asking whether the probability to reach a certain subset $G$ of goal states is at most some threshold~$p$---or to find a lower bound on $\E(\diamondsuit G)$, the expected reward (\eg runtime cost) accumulated before reaching~$G$.
In this paper, we specifically focus on probabilistic model checking (PMC) for (time-homogeneous) \emph{discrete-time} Markov chains (DTMCs) and Markov reward models (MRMs) given \emph{probabilistic reachability} properties like $\phi^G_{\leq p}$ or \emph{expected reward} queries for $\E(\diamondsuit G)$.

A DTMC is a stochastic process $\set{X_t}_{t \in T}$ over a discrete state space $S$ and (time) index set $T = \N$ that satisfies the Markov (memoryless) property.
An MRM adds a random variable mapping paths to real values.
DTMCs are classically defined via their transition kernel or probability matrix, that is, the one-step probabilities $\P(X_{i+1} = t \mid X_i = s)$ of moving from state $s$ to $t$ for all pairs of states, plus an initial state or initial distribution.
An MRM then additionally provides one-step reward values.
DTMCs and MRMs have been formalised in Isabelle/HOL in this view~\cite{Hol17b,Hol17,SA23}, starting from a Markov kernel upon which probability measures on sets of (infinite) paths or traces, expectations, etc.\ can be built.

When a non-probabilistic model checker finds that a property does not hold, its verdict is accompanied by a \emph{counterexample}, typically one path through the model.
In PMC, the notion of counterexample is more involved:
If $\smash[b]{\phi^G_{\leq p}}$ does not hold, then a counterexample could be a \emph{set} of paths, each of which reaches $G$ from the initial state, and which in total has a probability mass greater than~$p$.
These sets can quickly grow too large to be helpful in explaining the PMC result to the user, or to help them correct their model or system.
Various other high- and low-level notions of probabilistic counterexamples have thus been developed over the past two decades~\cite{AJJ25}---in essence, various ways of representing sets of paths in compact and explainable ways.
Among these, we are interested in techniques that \emph{abstract} the DTMCs we analyse in a \emph{refined} manner, and in particular the \emph{path abstraction} approach coined by \'{A}brah\'{a}m \etal in 2010~\cite{AJWKB10}.

Path abstraction can be seen as both a PMC algorithm and a counterexample refinement approach:
In the case of $\smash[b]{\phi^G_{\leq p}}$ not holding, the set of all paths that reach $G$ is a counterexample, and is what PMC algorithms usually (implicitly) compute.
Yet as a witness to the violation of the threshold, a much smaller set may be sufficient.
Path abstraction %
provides a \emph{heuristic} to \emph{structurally} find these smaller subsets of the (total) path sets, by iteratively \emph{abstracting} the DTMC over a sequence of subsets of $S - G$, %
the set of non-goal states.
Notably, performing path abstraction over $S - G$ directly is equivalent to PMC, and provides no refined diagnostic information; the contribution of path abstraction arises from the use of a well-chosen sequence of such abstraction sets.

\paragraph{State of the art.}
Path abstraction was originally formulated for DTMCs only~\cite{AJWKB10}, using their classical definition,
and tightly coupled to a decomposition into strongly connected components (SCCs) determining the sequence of abstraction sets.
In 2025, we proved that \emph{any} sequence of subsets of $S - G$ can be chosen to abstract the DTMC~\cite{HM25}. %
They also showed the two %
crucial correctness properties, which had not been explicitly addressed in~\cite{AJWKB10}:
(1)~Path-abstracting a DTMC yields another DTMC (\emph{well-definedness}~\cite[Lemma~2]{HM25}), and
(2)~path-abstracting a DTMC over a set of states $A$ and then over a set of states $B \supseteq A$ yields the same result as abstracting over $B$ straight away (\emph{monotonic absorption}~\cite[Thm.~3]{HM25}).

Both (1) and (2) were proven in a \emph{free monoid} view of DTMCs:
instead of the classical transition kernel $S\times S\to[0,1]$, they use one of the form $S^*\to[0,1]$ where $S^*$ is the \emph{free monoid} on $S$ (\ie the set of words over alphabet $S$),
enabling smoother correctness proofs (notably for monotonic absorption).
However, the proofs of~\cite{HM25} %
still fill ${\approx}\,{6}$ densely written pages in the paper's extended version~\cite[App.~A \& B]{HM25x} solely relying on the trust level of human-brain reasoning.

\paragraph{Our contributions.}
We put the work of~\cite{HM25} onto a formal, machine-checked foundation, and simultaneously extend it to MRMs:
(i)~We define MRMs in the free monoid view, and specify a new path abstraction algorithm for MRMs that preserves expected (cumulative) rewards.
(ii)~We formalise MRMs and DTMCs, and the two corresponding path abstraction algorithms, in Isabelle/HOL.
(iii)~For both algorithms, we mechanise well-definedness and monotonic absorption.

We thus provide machine-checked correctness proofs for counterexample refinement at the trust level of the Isabelle inference kernel.
For DTMC reachability, we build on the definitions and pen-and-paper proofs of~\cite{HM25x}; for expected rewards on MRMs, we provide the first free monoid-based definition and path abstraction algorithm, with completely new proofs made directly in Isabelle/HOL.
We also make explicit how path-abstracting over the non-goal states is equivalent to PMC, preserving reachability probabilities and expected rewards---which was stated but not proven in~\cite{HM25}.

Our work emphasises how an appropriate choice of repre\-sentation---based on an a-priori introspection to identify the minimal semantics/theory in which a problem (and its proofs) can be expressed---can significantly reduce the effort to apply interactive theorem proving (ITP).
Concretely, Isabelle/HOL's semantic framework around lists combines very naturally with the algebraic and list-recursive nature of the path abstraction algorithms.
The free monoid view additionally simplified our proof setup:
Where general formalisations like Hölzl's capture the classical probability space for a DTMC induced by the cylinder sets of all possible paths, we only look at the set of minimal paths (or words) \wrt the prefix order, thus taking a basis of that induced $\sigma$-algebra and circumventing having to reason about $\sigma$-algebras in the first place.
Path abstraction is then merely a discrete mathematics operation.
Where we need measure-theoretic reasoning, we draw on the HOL-Analysis library, at most having to prove a special case of Tonelli's theorem~\cite[Sec.~20.1]{RF18} for double infinite sums.
Connecting the representations, we show the correspondence between the classical and free monoid views in \Cref{sec:FreeMonoidView}.
Overall, the most conceptually involved (and time-consuming) part of our proofs was to show that reachability probabilities and expected rewards converge (\Cref{lem:prob-lang-finconf-startingpt-conv-to-0-1,thm:evt-conv}, see \Cref{remark:ConvergenceProofs}), which are essential conditions enabling the proofs of our main lemmas.
To the best of our knowledge, no similar statement has been shown using ITP in a semantics this lightweight for Markov models.

\paragraph{Outline.}
We recall, in \Cref{sec:preliminaries,sec:FreeMonoidView}, theory on free monoids and real analysis, then define MRMs and DTMCs in both the classical and free monoid views.
\Cref{sec:PathAbstractionAndPMC} introduces path abstraction and relates it to PMC.
We then present our formalisation and mechanisation of MRMs, path abstraction, and the corresponding Isabelle/HOL proofs in \Cref{sec:verified-path-abstraction}.

\section{Preliminaries}
\label{sec:preliminaries}
Let $\N=\{0,1,2,\ldots\}$ denote the set of natural numbers, and $\Ng=\N-\{0\}$ the set of positive integers.
For a set $S$, we let $2^S\equiv\set{S'\mid S'\subseteq S}$ denote its power set.
Given a function $f\colon X\to Y$ and a set $Y'\subseteq Y$, we denote by
$f^{-1}(Y')\equiv\set{x\in X\colon f(x)\in Y'}$
the pre-image of $Y'$ under $f$.
The universal set is denoted by $\U$ and is the equivalent of Isabelle's UNIV.

\subsection{Languages and the Free Monoid}
\label{subsec:lang-free-monoid}
Given a set $S$, we let $S^*$ denote the set of all words that have letters in \emph{alphabet} $S$, and let $\varepsilon$ denote the empty word.
$S^*$~is called the \emph{free monoid} on $S$, and is indeed a monoid with the empty word $\varepsilon$ as neutral element and concatenation of words as (an associative) binary operation on $S^*$.
Let $S^+$ denote the \emph{Kleene plus} of $S$: $S^+\equiv S^*-\set{\varepsilon}$.
The concatenation of two words $x$ and $y$ is denoted by $xy$.
Given a word $x$, we let $\sett(x)$ denote the smallest subset of $S$ containing the letters of $x$, \ie $\sett(x)=\bigcap_{A\subseteq S\colon x\in A^*}A$.
Note that $\sett(x)$ is finite.
Given $x,y\in S^*$ we say that $y$ is a \emph{prefix} of $x$, denoted by $y\leqslant x$, if there exists a word $z$ such that $x=yz$. Then, $\leqslant$ becomes a partial order on $S^*$, called the \emph{prefix order}. We say that $y$ is a \emph{factor} of $x$, denoted by $y\sqsubseteq x$, if there exist words $w,z$ such that $x=wyz$;
we let $|x|_y\equiv|\set{(w,z)\in S^*\times S^*\colon x=wyz}|$ denote the number of factor occurrences of $y$ in $x$.
For $x\in S^*$, we let $|x|$ denote the \emph{length} of $x$.
Note that $|x|=\sum_{a\in\sett(x)}|x|_a$.

Let $S^\omega=\set{s_0s_1s_2\ldots\colon \forall i\in\N.\,s_i\in S}$ denote the set of all infinite sequences over $S$. Note that one can concatenate a word over $S$ with an infinite sequence over $S$, and that a word over $S$ can be a prefix of an infinite sequence over $S$.

A set $L$ is called a \emph{language} over the alphabet $S$ if $L\subseteq S^*$, and we let $L_\perp\equiv\{x \in L\colon x'\in L\land x'\leqslant x\Longrightarrow x'=x\}$ denote the \emph{prefix-minimal part} of $L$. Note that $(L_\perp)_\perp=L_\perp$.

The concatenation of two languages $K,L\subseteq S^*$ is denoted by $KL$ and is defined by $KL\equiv\set{xy\colon x\in K\land y\in L}$.
Furthermore, the prefix order $\leqslant$ on $S^*$ extends to languages $K,L\subseteq S^*$: we write $K\leqslant L$ if $\forall x\in L.\,\exists y\in K.\,y\leqslant x$.
Note that $\leqslant$ is a pre-order on $2^{S^*}$ but not a partial order (any two distinct languages containing $\varepsilon$ precede each other \wrt $\leqslant$).

Whenever convenient, we interpret single symbols as words and words as singleton languages, and omit set brackets around words when referring to singleton languages.

Finally, given $x\in S^*$ we let $\star x$ denote the \emph{tail} of $x$, where $\star x\equiv\varepsilon$ if $x=\varepsilon$ and $\star x\equiv	x'$ if $x=ax'$ for some $a\in S$ and $x'\in S^*$.
The $\star$-operator extends to languages through $\star L\equiv\set{\star x\colon x\in L}$ (where $L\subseteq S^*$), and can hence also be used as the binary operation on $2^{S^*}$ where $K\star L=K(\star L)$.

\subsection{Infinite Sums and Real Analysis}
To reason with probability masses of infinite sets of paths, which are unordered infinite sums\footnote{With \emph{infinite sum} we refer to sums whose \emph{domains} are \emph{potentially} infinite.} of individual path probabilities, we default to the following definition of infinite sums, which is also the standard definition in \emph{HOL-Analysis}:

\begin{definition}
\label{def:infsum}
Let $f\colon A\to\R$ and $L\in\R$.
Then $f$ has \emph{infinite sum} $L$ \emph{on} $A$, denoted $\sum_{x\in A}f(x)=L$,
if
\begin{align}
&\forall\epsilon>0.\,\exists F.\,F\subseteq A\land|F|<\infty\label{eq:infsum-def}\\
&\land\forall F'.\,F\subseteq F'\subseteq A\land |F'|<\infty\Longrightarrow\left|\sum\nolimits_{x\in F'}f(x)-L\right|<\epsilon\notag.
\end{align}
\end{definition}
That is, $\sum_{x\in A}f(x)=L$ if relative to every positive error tolerance there exists a finite subset $F$ of $A$ such that we can approximate $L$ by summing $f$ over any finite subset of $A$ containing $F$.\!\footnote{Note that the classical definition of countable ordered sums $\sum_{i\in\N}x_i$ is subsumed by this (stronger) notion of infinite sum, as each partial sum is indexed by a set of the form $\set{0,\ldots,n}$ which is a finite subset of $\N$.}
Eq.~\eqref{eq:infsum-def} is trivially satisfied if $A$ itself is finite. %

\subsection{Markov Chains and Reward Models}
\label{subsec:MC-RM}

\begin{definition}
\label[definition]{def:MRM}
A (\emph{transition-based}) \emph{Markov reward model} (MRM) is a tuple $\M=(S, s_0,\P,R)$ with finite \emph{state space} $S$, \emph{initial state} $s_0\in S$, \emph{transition probability function} $\P\colon S\times S\to[0,1]$ with $\forall s\in S. \sum_{t\in S}\P(s,t)=1$, and \emph{reward function} $R\colon S\times S\to\R$ s.t.\ $\P(s,t)=0 \Rightarrow R(s,t)=0$.
\end{definition}

\begin{definition}
A \emph{discrete-time Markov chain} (DTMC) is a triple $M=(S,s_0,\P)$ such that $S,s_0,\P$ are the first three entries of an MRM. %
\end{definition}
Given an MRM $\M=(S,s_0,\P,R)$, we call $(S,s_0,\P)$ its \emph{underlying} or \emph{induced} DTMC.
A DTMC $M=(S,s_0,\P)$ further induces a directed graph $(S,E)$ with edge set $E=\set{(s,t)\in S\times S\colon\P(s,t)>0}$, called the \emph{underlying graph} of $M$, which can also be seen as the underlying graph of $\M$.

To a DTMC $M=(S,s_0,\P)$ we can associate the sets $M^+\subseteq S^+$ and $M^\omega\subseteq S^\omega$ of finite and infinite paths of its underlying graph, including paths that do not contain edges but just consist of a single state.
Both give rise to the following probability space on $M$ standard in the literature:

\begin{theorem}
\label{thm:sigma-algebra-MC}
Let $\F$ be the $\sigma$-algebra on $M^\omega$ generated\,\footnote{The $\sigma$-algebra $\sigma(\mathcal{A})$ on $\Omega$ generated by an $\mathcal{A}\subseteq 2^\Omega$ is by definition the smallest $\sigma$-algebra on $\Omega$ containing $\mathcal{A}$, which is well-defined as the intersection of $\sigma$-algebras on a common set is again a $\sigma$-algebra.} by the set $\set{xS^\omega\cap M^\omega\colon x\in M^+}\subseteq2^{S^\omega}$.
Then, there exists a unique function $\mu:\F\to[0,1]$ satisfying, for all $x\in M^+$,
\begin{equation}
\label{eq:measure-cases}
\mu(xS^\omega\cap M^\omega)=\begin{cases}
\P(x),&\text{if }x=x_0\ldots x_n\land x_0=s_0\land n>0,\\
1,&\text{if }x=s_0,\\
0,&\text{otherwise,}
\end{cases}
\end{equation}
such that $(M^\omega,\F,\mu)$ is a probability space,
where $\P(x)$ is short for
$$\prod_{i=0}^{n-1}\P(x_i,x_{i+1})$$
in the first case of~\eqref{eq:measure-cases}.
\end{theorem}
\begin{proof}
Argued in (the context of)~\cite[Def.~10.10]{BK08}.
\end{proof}

\begin{definition}
\label{def:reach-prob}
Given a set $G\subseteq S$ of \emph{goal states}, the unbounded \emph{reachability probability} of reaching $G$ in $M$ is
\begin{equation}
\label{eq:reach-prob}
\P(\diamondsuit G)\equiv\mu(M^\omega-(S-G)^\omega).
\end{equation}
\end{definition}

\begin{convention}
\label{conv:initial-state-no-goal-state}
We assume throughout that any set of goal states does \emph{not} contain the initial state.
We do this because in the case $s_0\in G$ we trivially have $\P(\diamondsuit G)=1$, and there would be nothing substantial left to compute or to reason with.
\end{convention}

\begin{remark}
Note that $M^\omega-(S-G)^\omega$ is the set of all paths in $M$ that contain a state in $G$, and that
\begin{equation}
\label{eq:reachability-union}
M^\omega-(S-G)^\omega=\biguplus\nolimits_{x\in M^+\cap (S-G)^*G}xS^\omega\cap M^\omega
\end{equation}
is a disjoint\footnote{The union in~\eqref{eq:reachability-union} is disjoint because no two distinct elements of $(S-G)^*G$ are prefixes of each other.} countable union of sets that are measurable by definition of $\F$ in \Cref{thm:sigma-algebra-MC} and hence is itself measurable. As such,~\eqref{eq:reach-prob} is well-defined.
\end{remark}

\begin{definition}
\label{def:exp-rew}
Let $\M=(S,s_0,\P,R)$ be an MRM. Then, given a set $G\subseteq S$ of goal states, the \emph{expected reward} of reaching $G$ in $\M$ is defined by
\begin{equation}
\label{eq:def-exp-rew}
\E(\diamondsuit G)\equiv \E[X]=\int_{M^\omega} X(x)\,d\mu(x)
\end{equation}
where $\E[X]$ is the expected value of the random variable $X:M^\omega\to\R$ defined by, where we write $x=x_0x_1\ldots$,
\begin{equation}
\label{eq:exp-rew}
X(x)\equiv\begin{cases}
\sum_{i=0}^{n-1}R(x_i,x_{i+1})&\begin{aligned}
&\text{if }x\in M^\omega-(S-G)^\omega\\[-4pt]
&\land\, x_0\ldots x_{n-1}\in (S-G)^* \land x_n\in G,
\end{aligned}\\
0&\text{otherwise.}
\end{cases}
\end{equation}
\end{definition}

\begin{remark}
Note that $n$ in~\eqref{eq:exp-rew} exists as an element of $\N$ and is unique if $x\in M^\omega-(S-G)^\omega$, with the conventions that the sum and the second clause of the condition of the first case of~\eqref{eq:exp-rew} are zero and vacuous, respectively, in the case $n=0$.
Note furthermore that $X$ in \Cref{def:exp-rew} is indeed a random variable, as the image $X(M^\omega)\subseteq\R$ of $M^\omega$ under $X$ is countable and the inverse image of every real number under $X$ is a countable union of the basis sets $xS^\omega\cap M^\omega$ (where $x\in M^+$) of the $\sigma$-algebra $\F$ from \Cref{thm:sigma-algebra-MC},
together with possibly $M^\omega\cap (S-G)^\omega$,
whose measurability is seen through
\begin{equation}
M^\omega\cap(S-G)^\omega=\bigcap\nolimits_{n\in\N}\bigcup\nolimits_{x\in M^+\cap (S-G)^{n+1}}xS^\omega\cap M^\omega,
\end{equation}
which is a countable intersection of countable unions of basis sets $xS^\omega\cap M^\omega$ of the $\sigma$-algebra $\F$ from \Cref{thm:sigma-algebra-MC}, hence measurable.
\end{remark}

\subsection{Markov Models in Isabelle/HOL}
The notions of MRMs and DTMCs have been formalised in Isabelle, most notably by H\"{o}lzl \etal~\cite{Hol17},
who formalise a DTMC as a transition kernel mapping a state to a discrete probability distribution indexed by its successor states.
In their formalisation, reachability probabilities are defined as the measure of the event in the trace space that eventually the goal set is reached.
The trace space's sample space is the set of all infinite sequences over the state space---hence possibly including those of zero probability---with a $\sigma$-algebra and measure generated and defined analogously to those in \Cref{thm:sigma-algebra-MC}.
Expected rewards in~\cite{Hol17} are expressed as least fixpoints of a functional that sums up state pair rewards until the set of goal states is reached, and hence only admit non-negative rewards; more recent work extends to rewards with arbitrary real values~\cite{SA23}. %

\section{A Free Monoid View on Markov Models}
\label{sec:FreeMonoidView}

For reasoning about reachability probabilities in DTMCs with finite state spaces, a lightweight semantics that does not involve measure theory suffices~\cite{HM25}; the same holds for expected rewards in MRMs.
The quantities of interest $\P(\diamondsuit G)$ in~\eqref{eq:reach-prob} and $\E(\diamondsuit G)$ in~\eqref{eq:def-exp-rew} can be expressed as infinite sums over finite paths.
In the setting of \Cref{thm:sigma-algebra-MC} and \Cref{def:reach-prob,def:exp-rew}, this is reflected in the following result.

\begin{theorem}
\label{thm:reach-prob-exp-rew-infsums}
For an MRM $\M=(S,s_0,\P,R)$, we have
\begin{equation}
\label{eq:reach-prob-exp-rew-infsums}
\begin{split}
\P(\diamondsuit G)&=\sum\nolimits_{x\in s_0(S-G)^*G}\P(x),\\
\E(\diamondsuit G)&=\sum\nolimits_{x\in s_0(S-G)^*G}\P(x)R(x),\\
\end{split}
\end{equation}
where $\P(x)$ is short for $\prod_{i=0}^{n-1}\P(x_i,x_{i+1})$ for $x=x_0\ldots x_n$ and $n\in\N^+$, as written earlier in \Cref{thm:sigma-algebra-MC}, and, analogously, $R(x)$ is short for $\sum_{i=0}^{n-1}R(x_i,x_{i+1})$.
\end{theorem}

\begin{proof}[Proof of \Cref{thm:reach-prob-exp-rew-infsums}]
For $\P(\diamondsuit G)$, we use~\eqref{eq:reach-prob} and~\eqref{eq:reachability-union} and the countable additivity rule of measures to deduce that $\P(\diamondsuit G)=\sum_{x\in M^+\cap(S-G)^*G}\mu(xS^\omega\cap M^\omega)$.
Note that $\mu(xS^\omega\cap M^\omega)=\P(x)$ for each $x\in M^+\cap (S-G)^*G$ by~\eqref{eq:measure-cases},
and hence $\P(\diamondsuit G)=\sum_{x\in M^+\cap(S-G)^*G}\P(x)$.
Since $\mu(xS^\omega\cap M^\omega)=0$ if $x\in S^+$ does not start in $s_0$, we can replace the domain $M^+\cap(S-G)^*G$ by $M^+\cap(S-G)^*G\cap s_0S^*$.
Since $s_0\notin G$ by \Cref{conv:initial-state-no-goal-state}, we have $(S-G)^*G\cap s_0S^*=s_0(S-G)^*G$.
Then, we obtain $\P(\diamondsuit G)=\sum_{x\in M^+\cap s_0(S-G)^*G}\P(x)$, and, given that $\P(x)=0$ for all $x\in S^+-M^+$ by definition of $M^+$, $\P(\diamondsuit G)=\sum_{x\in s_0(S-G)^*G}\P(x)$, and we have verified the first line of~\eqref{eq:reach-prob-exp-rew-infsums}.

For $\E(\diamondsuit G)$, starting from~\eqref{eq:exp-rew}, we note that the random variable $X$ encoding $R$ on finite paths is constant on each basis set $xS^\omega\cap M^\omega$ ($x\in M^+$) of the $\sigma$-algebra of the probability space on $M$ (it is constant with value $R(x)$ or with value $0$), and hence its expected value is the sum of all involved path probabilities multiplied by their respective cumulative rewards.
By this argument, given that we have established the first line of~\eqref{eq:reach-prob-exp-rew-infsums}, its second line follows.
\end{proof}
We later show that the two sums of \eqref{eq:reach-prob-exp-rew-infsums} in fact converge:
this is \Cref{lem:prob-lang-finconf-startingpt-conv-to-0-1} for probabilities, and follows from \Cref{thm:evt-conv} for expected rewards.

So far, for convenience, we have used $\P(x)$ and $R(x)$ as abbreviations of the probability and (cumulative) reward of a path $x$.
Since we need to reason a lot with path probabilities and cumulative rewards, we adopt the free monoid view of~\cite{HM25}, which defines DTMCs via probabilities of all non-empty words over the state space, and extend it to MRMs.
Since we have defined DTMCs as structures underlying MRMs in \Cref{def:MRM}, we only need to rework this exact definition.

\begin{definition}[MRM in a free monoid view]
\label{def:MRM-free-monoid}
A Markov reward model (MRM) is a quadruple $\M=(S,s_0,\P,R)$ where $S$ is a finite set called the \emph{state space} of $\M$; $s_0\in S$ is the \emph{initial state} of $\M$; $\P:S^+\to[0,1]$ is a function satisfying, for all $s\in S$, $\sum_{t\in S}\P(st)=1$ and $\P(s)=1$ and $x,y\in S^*\Rightarrow\P(xsy)=\P(xs)\P(sy)$; and $R:S^+\to\R$ is a function satisfying $R(x)=0$ whenever $|x|=1\lor\P(x)=0$ and, for all $s\in S$ and $x,y\in S^*$ for which $\P(xsy)>0$, $R(xsy)=R(xs)+R(sy)$.
\end{definition}
The above definition captures the fact that the function $\P$ must mimic a probability distribution per state indexed by its successor states via two-state sequences; the Markov property is captured by the rule that we can break down a path at any point such that the probability of the whole path equals the product of the probabilities of the two segments.
Furthermore, the cumulative reward function's \emph{cumulative} property---namely, on a path with non-zero probability, its cumulative reward is the sum of the rewards of the transitions on the path---is captured by a similar rule of breaking a path into two segments; finally, (cumulative) rewards on paths that consist of a single state or have probability zero must be zero as well.

\begin{remark}
The correspondence between MRMs in the classical view of \Cref{def:MRM} and MRMs in the free monoid view of \Cref{def:MRM-free-monoid} can be seen as follows.
By the rule in \Cref{def:MRM-free-monoid} that $\P(xsy)=\P(xs)\P(sy)$ for all $s\in S$ and $x,y\in S^*$, the function $\P$ is fully determined by its values on two-state sequences.
A similar argument applies to the reward function.
\end{remark}

\begin{convention}
\label{conv:substochastic}
Within PMC, the condition $\sum_{t\in S}\P(st)=1$ for all $s\in S$ in \Cref{def:MRM-free-monoid} is often too strict and hence relaxed to $\sum_{t\in S}\P(st)\leqslant1$ for all $s\in S$.
From now on, we adopt this convention.
\end{convention}
In the free monoid view of MRMs, $\P(\diamondsuit G)$ and $\E(\diamondsuit G)$ are defined directly by \eqref{eq:reach-prob-exp-rew-infsums}, which \Cref{thm:reach-prob-exp-rew-infsums} relates back to the original \Cref{def:reach-prob,def:exp-rew} in the classical view.
From now on, we work in the free monoid view exclusively.

\section{Path Abstraction and PMC}
\label{sec:PathAbstractionAndPMC}

In this section, we illustrate path abstraction with an example, restate its definition for DTMCs, and extend it to MRMs.
We then show that abstracting over all non-goal states yields reachability probabilities and expected rewards exactly, while intermediate abstractions can already provide bounds that suffice to decide whether a threshold is met.
The mechanised proofs are included in our supplementary material (file \texttt{supplementary-material.zip}), and we refer to the specific files for each theorem from now on.

\subsection{Motivating Example}
\label{sec:motivation}

Let us motivate our interest in the path abstraction technique with the example program in~\Cref{fig:prob-prog}, which contains a fair coin flip.

\begin{figure}[t]
\centering
\begin{minipage}[b]{0.365\textwidth}
\centering
\begin{lstlisting}[mathescape=true, emph={while,if,else,end},
  emphstyle=\bfseries, label=lstlisting:prob-prog-labelled]
$x\leftarrow0$, $y\leftarrow0$
while $x<2$:
  if $rand()\leqslant\frac{1}{2}$: $x\leftarrow x+1$
  else: $y\leftarrow -y+1$
$y\leftarrow 2\cdot(-y+1)^{-1}$
\end{lstlisting}
\caption{A probabilistic program}
\label{fig:prob-prog}
\end{minipage}%
\hfill
\begin{minipage}[b]{0.625\textwidth}
\centering
\begin{tikzpicture}
\draw
(2, 1) node(ghost){}
(3, 1) node[squared node, inner xsep=2pt](s00){$0,0$}
(5, 1) node[squared node, inner xsep=2pt](s10){$1,0$}
(7, 1) node[squared node, inner xsep=2pt](s20){$2,0$}
(9, 1) node[squared node, inner xsep=2pt](s22){$2,2$}
(3, -1) node[squared node, inner xsep=2pt](s01){$0,1$}
(5, -1) node[squared node, inner xsep=2pt](s11){$1,1$}
(7, -1) node[squared node, inner xsep=2pt](s21){$2,1$}
(9, -1) node[squared node, inner xsep=2pt](s2inf){$2,\infty$};
\draw
(4, 1.33) node{$\frac{1}{2};1$}
(6, 1.33) node{$\frac{1}{2};1$}
(8, 1.2) node{$1;5$}
(4, -0.67) node{$\frac{1}{2};1$}
(6, -0.67) node{$\frac{1}{2};1$}
(8, -0.8) node{$1;5$}
(2.35,0.5) node{$\frac{1}{2};2$}
(4.35,0.5) node{$\frac{1}{2};2$}
(3.7,0) node{$\frac{1}{2};2$}
(5.7,0) node{$\frac{1}{2};2$};
\draw[-stealth] (ghost) -- (s00);
\draw[-stealth] (s00) -- (s10);
\draw[-stealth] (s10) -- (s20);
\draw[-stealth] (s01) -- (s11);
\draw[-stealth] (s11) -- (s21);
\draw[-stealth] (s20) -- (s22);
\draw[-stealth] (s21) -- (s2inf);
\draw[-stealth] (s00) to[out = -60, in = 60] (s01);
\draw[-stealth] (s01) to[out = 120, in = -120] (s00);
\draw[-stealth] (s10) to[out = -60, in = 60] (s11);
\draw[-stealth] (s11) to[out = 120, in = -120] (s10);
\end{tikzpicture}
\caption{The MRM of \Cref{fig:prob-prog} with computation times}
\label{fig:prob-prog-MRM}
\end{minipage}
\end{figure}

We can model the behaviour of this program as a DTMC by composing a state space of all possible values the pair $(x,y)$ of variables can take, where an undefined division is indicated with a $+\infty$-sign:
\begin{equation}
S=\set{(0,0),(0,1),(1,0),(1,1),(2,0),(2,1),(2,2),(2,\infty)}.
\end{equation}
Here, the initial state is $(0,0)$.
Additionally, we would like to capture the (expected) \emph{time} it takes to reach a terminal state. Suppose that a $+1$-increment, doubling, and additive inversion take $1$ time unit each, and multiplicative inversion takes $2$ time units.
We can now model the program in \Cref{fig:prob-prog} with these computation time costs as a transition-based reward function %
into a full MRM, shown in~\Cref{fig:prob-prog-MRM}.

\begin{figure}
\centering
\begin{tikzpicture}
\draw
(1.5, 1) node(ghost){}
(2.5, 1) node[squared node, inner xsep=2pt](s00){$0,0$}
(5, 1) node[squared node, inner xsep=2pt, fill=gray!25](s10){$1,0$}
(7, 1) node[squared node, inner xsep=2pt](s20){$2,0$}
(9, 1) node[squared node, inner xsep=2pt](s22){$2,2$}
(2.5, -1) node[squared node, inner xsep=2pt](s01){$0,1$}
(5, -1) node[squared node, inner xsep=2pt, fill=gray!25](s11){$1,1$}
(7, -1) node[squared node, inner xsep=2pt](s21){$2,1$}
(9, -1) node[squared node, inner xsep=2pt](s2inf){$2,\infty$};
\draw
(3.5, 1.33) node{$\frac{1}{2};1$}
(6, 1.33) node{$\frac{1}{2};1$}
(8, 1.2) node{$1;5$}
(3.5, -0.67) node{$\frac{1}{2};1$}
(6, -0.67) node{$\frac{1}{2};1$}
(8, -0.8) node{$1;5$}
(1.85,0.5) node{$\frac{1}{2};2$}
(4.35,0.5) node{$\frac{1}{2};2$}
(3.2,0) node{$\frac{1}{2};2$}
(5.7,0) node{$\frac{1}{2};2$};
\draw[-stealth] (ghost) -- (s00);
\draw[-stealth] (s00) -- (s10);
\draw[-stealth] (s10) -- (s20);
\draw[-stealth] (s01) -- (s11);
\draw[-stealth] (s11) -- (s21);
\draw[-stealth] (s20) -- (s22);
\draw[-stealth] (s21) -- (s2inf);
\draw[-stealth] (s00) to[out = -60, in = 60] (s01);
\draw[-stealth] (s01) to[out = 120, in = -120] (s00);
\draw[-stealth] (s10) to[out = -60, in = 60] (s11);
\draw[-stealth] (s11) to[out = 120, in = -120] (s10);
\draw[thick, densely dotted] (3.9,1.7) -- (3.9,-1.7);
\draw[thick, densely dotted] (3.9,-1.7) -- (7.6,-1.7);
\draw[thick, densely dotted] (7.6,-1.7) -- (7.6,1.7);
\draw[thick, densely dotted] (7.6,1.7) -- (3.9,1.7);
\end{tikzpicture}
\caption{The MRM of \Cref{fig:prob-prog} with sub-MRM induced by $A=\set{(1,0),(1,1)}$}
\label{fig:prob-prog-MRM-abstr-selection}
\end{figure}
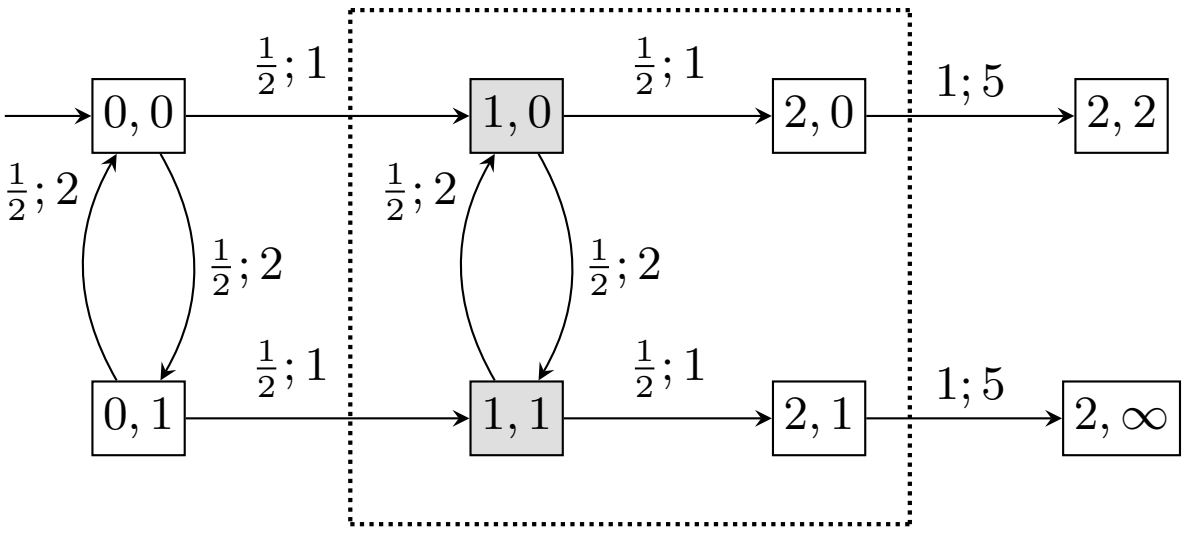

Now, let us illustrate the path abstraction technique~\cite{AJWKB10,HM25}.
Suppose we only want to run our program if $\P(\diamondsuit(2,\infty))\leqslant15\%$.
Rather than computing this probability exactly, we want to find a lower bound that exceeds $15\%$ to conclude that we shall not run our program.
To this end, let us consider the sub-MRM ``induced'' by $A=\set{(1,0),(1,1)}$, with state space $A$ together with the states outside of $A$ that can be reached from $A$ within one step as terminal states.
\Cref{fig:prob-prog-MRM-abstr-selection} shows $A$ in grey and the sub-MRM within dotted lines.

\begin{figure}[t]
\centering
\begin{tikzpicture}
\draw
(1.5, 1) node(ghost){}
(2.5, 1) node[squared node, inner xsep=2pt](s00){$0,0$}
(4.5, 1) node[squared node, inner xsep=2pt](s10){$1,0$}
(7, 1) node[squared node, inner xsep=2pt](s20){$2,0$}
(9, 1) node[squared node, inner xsep=2pt](s22){$2,2$}
(2.5, -1) node[squared node, inner xsep=2pt](s01){$0,1$}
(4.5, -1) node[squared node, inner xsep=2pt](s11){$1,1$}
(7, -1) node[squared node, inner xsep=2pt](s21){$2,1$}
(9, -1) node[squared node, inner xsep=2pt](s2inf){$2,\infty$};
\draw
(3.5, 1.33) node{$\frac{1}{2};1$}
(6.15, 1.33) node{$\frac{2}{3};\frac{7}{3}$}
(8, 1.2) node{$1;5$}
(3.5, -0.67) node{$\frac{1}{2};1$}
(6.15, -0.67) node{$\frac{2}{3};\frac{7}{3}$}
(8, -0.8) node{$1;5$}
(1.85,0.5) node{$\frac{1}{2};2$}
(3.2,0) node{$\frac{1}{2};2$}
(5,0.4) node{$\frac{1}{3};\frac{13}{3}$}
(5,-0.4) node{$\frac{1}{3};\frac{13}{3}$};
\draw[-stealth] (ghost) -- (s00);
\draw[-stealth] (s00) -- (s10);
\draw[-stealth] (s10) -- (s20);
\draw[-stealth] (s01) -- (s11);
\draw[-stealth] (s11) -- (s21);
\draw[-stealth] (s20) -- (s22);
\draw[-stealth] (s21) -- (s2inf);
\draw[-stealth] (s00) to[out = -60, in = 60] (s01);
\draw[-stealth] (s01) to[out = 120, in = -120] (s00);
\draw[-stealth] (s10) to[bend left = 20] (s21);
\draw[-stealth] (s11) to[bend left = -20] (s20.south west);
\end{tikzpicture}
\caption{The MRM abstracted over $A$ into a new MRM}
\label{fig:prob-prog-MRM-abstr}
\end{figure}
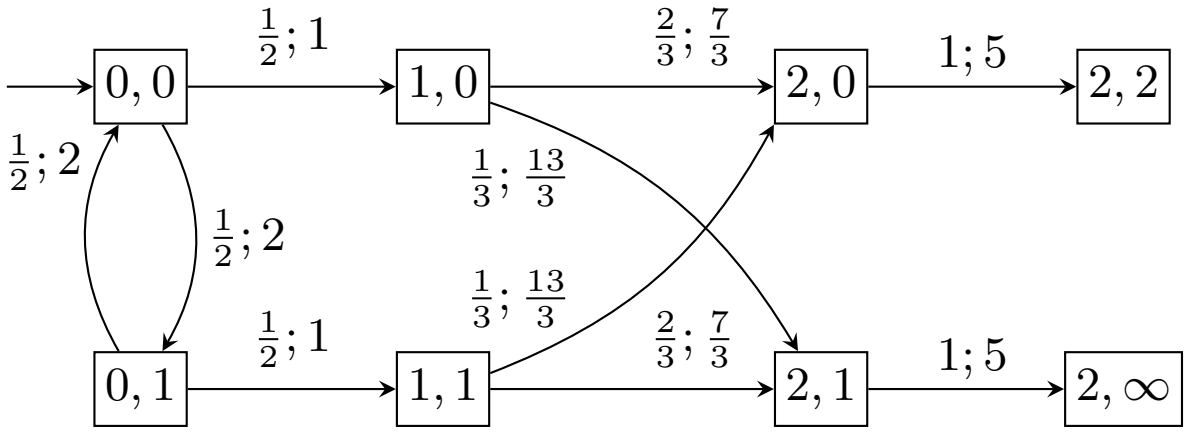

Now, in this sub-MRM, we reroute the total reachability probabilities and expected rewards (divided by the respective reachability probabilities) onto new transitions from the states of $A$ from which one can reach an outside state (in this case all of $A$) to those outside states.
The result is shown in \Cref{fig:prob-prog-MRM-abstr}.
Here, the abstract path $x=(0,0)(1,0)(2,1)(2,\infty)$ has probability $\frac{1}{6}$, which is greater than $15\%$.
Since path abstraction preserves\footnote{Under the common assumption in PMC that there are no transitions of positive probability between the goal states.} reachability probabilities, we have $\P(\diamondsuit(2,\infty))=\P'(\diamondsuit(2,\infty))$ where $\P$ and $\P'$ are the transition probability functions of the MRMs of \Cref{fig:prob-prog-MRM} and \Cref{fig:prob-prog-MRM-abstr}, respectively. Clearly, $\P'(x)=\frac{1}{6}$ contributes to the sum constituting $\P'(\diamondsuit(2,\infty))$---see \Cref{thm:reach-prob-exp-rew-infsums}---and thus $\P(\diamondsuit(2,\infty))=\P'(\diamondsuit(2,\infty))\geqslant\frac{1}{6}>15\%$, and we do not run our program.
Note that $\P(\diamondsuit(2,\infty))>15\%$ was not deducible in the analogous manner from a single path in the original MRM in \Cref{fig:prob-prog-MRM}, as the largest probability of a path from $(0,0)$ to $(2,\infty)$ is $\frac{1}{8}=12.5\%<15\%$.

We can also obtain information about other quantities of interest.
For example, summing probability times reward over the two paths $(0,0)(1,0)(2,0)(2,2)$ and $(0,0)(0,1)(1,1)(2,1)(2,\infty)$ in \Cref{fig:prob-prog-MRM-abstr} yields
\begin{equation}
\label{eq:exp-val-lower-bnd}
\frac{1}{3}\cdot(1+\frac{7}{3}+5)+\frac{1}{6}\cdot(2+1+\frac{7}{3}+5)
=\frac{9}{2}
\end{equation}
which serves as a lower bound of $\E(\diamondsuit\set{(2,2),(2,\infty)})$.

\subsection{The Path Abstraction Algorithms}
We now restate the definition of path abstraction for DTMCs~\cite{AJWKB10} in the free monoid view~\cite[Def.~6]{HM25}.
To this end, we fix a DTMC $M=(S,s_0,\P)$ in the free monoid view, as defined in \Cref{def:MRM-free-monoid}, with the relaxed substochasticity convention~\Cref{conv:substochastic}.

\begin{definition}
\label{def:prob-lang}
Given $L\subseteq S^+$, let $\P(L)$ denote the infinite sum %
\begin{equation}
\label{eq:prob-lang}
\P(L)\equiv\sum\nolimits_{x\in L_\perp}\P(x),
\end{equation}
where we recall that $L_\perp$ is the set of minimal elements of $L$ with respect to the prefix order~$\leqslant$.
\end{definition}
In~\eqref{eq:prob-lang} we mostly focus on those $L$ of the form $L=aL'$ with $a\in S$ and $L'\subseteq S^*$.
Then, the interpretation of $\P(L)$ is that it is the probability that an infinite trajectory starting in state $a$ initially moves along a path in $L$.
Indeed: this probability is equal to the measure of the union of all basis sets $xS^\omega\cap M^\omega$ where $x$ ranges over $L$ (where the $\sigma$-algebra involved is taken \wrt initial state $a$), which can be written as the disjoint union of all $xS^\omega\cap M^\omega$ where $x$ ranges over the \emph{prefix-minimal} part $L_\perp$ of $L$---\ie all minimal elements of $L$ \wrt the prefix order---because $x\leqslant x'$ always implies $x'S^\omega\cap M^\omega\subseteq xS^\omega\cap M^\omega$.
This union is disjoint because no two distinct elements of $L_\perp$ are prefixes of one another; since the probability of each $xS^\omega\cap M^\omega$ is $\P(x)$ we can indeed conclude that $\P(L)$ is the probability that an infinite trajectory starting in $a$ first moves along a path in $L$.

\begin{remark}
The reason for this generality of~\eqref{eq:prob-lang}, as opposed to specifying $\P(L)$ only relative to languages $L$ of the form $L=s_0L'$ for the fixed initial state $s_0$ of the DTMC, is that we are analysing ``sub-chains'' and ``sub-MRMs'' within larger systems and need flexibility around starting states.
This is completely analogous to how \'{A}brah\'{a}m \etal~\cite{AJWKB10} deal with probabilities of path sets represented by a set of finite prefixes.
\end{remark}

\begin{definition}
\label{def:inp}
For a DTMC $M=(S,s_0,\P)$ and a set of states $A\subseteq S$,
let
\begin{equation}
\inp(M,A)\equiv\set{s\in S\colon s=s_0\lor \P((S-A)s)>0}
\end{equation}
denote the set of \emph{input states} of $A$.
\end{definition}
In other words, $\inp(M,A)$ is the set of states in $A$ that either coincide with the initial state of $M$ or are reachable from outside of $A$ in one step.
For example, the input set of $\set{(1,0),(1,1)}$ in the DTMC underlying the MRM in \Cref{fig:prob-prog-MRM-abstr-selection} is simply $\set{(1,0),(1,1)}$ again; had we chosen the set of all states except for $(2,2)$ and $(2,\infty)$, then its input set would simply have been $\set{(0,0)}$.

Before we define the path abstraction algorithm, we need a compact method, enabling a smooth mechanisation in Isabelle, to represent languages that are of the form $L_1L_2\ldots L_n$ where $L_i=a_iA^*$ for some $a_i\in A$ for every odd $i$ and $L_i=b_i$ for some $b_i\notin A$ for every even $i$, or the other way around, where $A\subseteq S$, that we need to express probability masses in, for the path abstraction algorithm.

\begin{definition}
\label{def:red}
Define $\red:2^S\times S^*\to S^*$ by, for each $A\subseteq S$, defining $\red(A):S^*\to S^*\colon x\mapsto\red(A,x)$ inductively with two base cases.
Thus, let $A\subseteq S$. We set $\red(A)(\varepsilon)\equiv\varepsilon$; for all $a\in S$ (interpreted as $1$-letter words), we set $\red(A)(a)\equiv a$; for all $a,b\in S$ and $x\in S^*$, we set
\begin{equation}
\red(A)(abx)\equiv\begin{cases}
\red(A)(ax),&\text{if }a\in A\land b\in A,\\
a(\red(A)(bx)),&\text{otherwise.}
\end{cases}
\end{equation}
\end{definition}
The above recursive definition terminates, as the second argument decreases in length upon recursion.
Intuitively, $\red(A)$ replaces, when applied to $x$, all its maximal factors over $A$ by their respective first letters.
The function $\red$ comes in handy in representing languages of the form $L_1L_2\ldots L_n$ where $L_i=a_iA^*$ for some $a_i\in A$ for every odd $i$ and $L_i=b_i$ for some $b_i\notin A$ for every even $i$, or the other way around, where $A\subseteq S$:
they are equal to, respectively,
$$
\begin{aligned}
&(\red(A)^{-1}(a_1b_2\ldots(a\mid b)_n))_\perp\\
\text{and}\quad&(\red(A)^{-1}(b_1a_2\ldots(a\mid b)_n))_\perp.
\end{aligned}
$$

\begin{definition}
\label{def:abstr}
Let $M=(S,s_0,\P)$ be a DTMC.
For $A\subseteq S$, set $\abstr(A,M)\equiv(S,s_0,\P')$ where $\P':S^+\to[0,1]$ is defined\footnote{In~\cite[App.~B]{HM25x} it is proven that this $\P'$ indeed maps into $[0,1]$.} by $\P'(s)\equiv1$ for all $s\in S$ (interpreted as $1$-letter words here), and, for all $x\in S^+$ with $|x|\geqslant 2$,
\begin{equation}
\label{eq:abstr}
\P'(x)\equiv\begin{cases}
\P(\red(A)^{-1}(x)),&\text{if }x\in(S-(A-\inp(M,A)))^*,\\
0,&\text{otherwise.}
\end{cases}
\end{equation}
In obtaining $\abstr(A,M)$ from $M$, the set $A$ is called the \emph{abstraction set}.
\end{definition}
Path abstraction yields a DTMC again~\cite[Lemma~2]{HM25}:
\begin{theorem}[\texttt{Abstr.thy/abstr{\_}well{\_}defined}]
\label{thm:PA-MC-well-defined}
If $M=(S,s_0,\P)$ is a DTMC in the free monoid view of \Cref{def:MRM-free-monoid}, so is $\abstr(A,M)$, for any $A\subseteq S$.
\end{theorem}
To understand the new DTMC $(S,s_0,\P')$, we can examine the values $\P'$ takes on two-state sequences, as they fully determine $\P'$, because of \Cref{thm:PA-MC-well-defined}.
Thus let $s,t\in S$.
From \Cref{def:red} we establish that
\begin{equation}
\label{eq:red-char}
\red(A)^{-1}(st)=
\begin{cases}
\varnothing,&\text{if }s\in A\land t\in A,\\
sA^*t,&\text{if }s\in A\land t\notin A,\\
stA^*,&\text{if }s\notin A\land t\in A,\\
st,&\text{if }s\notin A\land t\notin A,
\end{cases}
\end{equation}
and thus applying $\perp$ to these languages gives $\varnothing$, $sA^*t$, $st$, and $st$, respectively.
Taking into account \Cref{def:prob-lang} and~\eqref{eq:abstr}, and writing $\varphi=s\in A-\inp(M,A)\lor t\in A-\inp(M,A)$ for convenience, we thus obtain
\begin{equation}
\label{eq:char-P'(st)}
\P'(st)=\begin{cases}
0,&\text{if }s\in A\land t\in A\lor\varphi,\\
\P(sA^*t),&\text{if }s\in A\land t\notin A\land\neg\varphi,\\
\P(st),&\text{otherwise.}
\end{cases}
\end{equation}
In other words, transitions (with positive probabilities) between states in $A$ are removed, and transitions from or to states in $A$ that are not part of the set $\inp(M,A)$ of input states of $A$ are removed as well; transitions from outside of $A$ to $\inp(M,A)$ are left unaltered, and transitions from $\inp(M,A)$ to the outside of $A$ are replaced by the total probability of starting in a state in $\inp(M,A)$, moving through $A$, and then ending up in a state outside of $A$ reachable from $A$ in one step.
Intuitively, we are modifying the sub-chain induced by $A$, \ie the chain with state space $A\cup\set{s\in S-A\colon\P(As)>0}$, into a chain where we remove all of its transitions but introduce new transitions from $\inp(M,A)$ to $\set{s\in S-A\colon\P(As)>0}$ with probabilities equal to the respective reachability probabilities within that sub-chain.

Proving or mechanising \Cref{thm:PA-MC-well-defined} is more difficult than it would be in the classical view, but in return, theorems about applying path abstraction \emph{multiple times} become easier to prove, such as \emph{monotonic absorption}~\cite[Thm.~3]{HM25}:

\begin{theorem}[\texttt{Abstr.thy/monotonic{\_}absorption}]
\label{thm:PA-MC-monabs}
Let $M=(S,s_0,\P)$ be a free-monoid DTMC as per \Cref{def:MRM-free-monoid} and ~\Cref{conv:substochastic}.
If $A\subseteq B\subseteq S$, then we have $\abstr(B,\abstr(A,M))=\abstr(B,M)$.
\end{theorem}

\begin{remark}
Since adding or removing isolated states (states that have zero incoming and zero outgoing probability) from an abstraction set has no effect on the abstraction, we may at each abstraction stage simply remove them \emph{or} add them.
This means in practice that (1) along a sequence of path abstractions, we can shrink the state space each time to the set of non-isolated states, reducing state space complexity along the way, and (2) as a consequence of (1), the SCC heuristic of~\cite{AJWKB10} for choosing sequences of abstraction sets is \emph{sound}, in the sense that it preserves total reachability probabilities and expected rewards to reach a set of goal states.
Since defining and verifying correctness of a version of path abstraction that does incorporate shrinkage of state spaces in line with (1) adds little conceptual or proof engineering complexity, we instead focus on the pure monotonic absorption properties, \Cref{thm:PA-MC-monabs} and \Cref{thm:PA-MRM-monabs}, themselves, and on developing the machinery to verify these properties. Then, the machinery can straightforwardly be extended to match concrete heuristics.
\end{remark}

Now, we extend \Cref{def:abstr} to MRMs, aiming to preserve \emph{expected rewards}.

\begin{definition}
\label{def:abstr-MRM}
Let $\M=(S,s_0,\P,R)$ be an MRM, with $M=(S,s_0,\P)$. Let $A\subseteq S$.
Writing $(S,s_0,\P')=\abstr(A,M)$ as in \Cref{def:abstr}, set $\abstr(A,\M)\equiv(S,s_0,\P',R')$ where $R':S^+\to\R$ is defined by, for $x\in S^+$,
\begin{equation}
\label{eq:abstr-MRM}
R'(x)\equiv\begin{cases}
\frac{1}{\P'(x)}\sum_{y\in(\red(A)^{-1}(x))_\perp}\P(y)R(y),&\begin{aligned}\text{if }|x|>1\land\\\P'(x)>0,\end{aligned}\\
0,&\text{otherwise.}
\end{cases}
\end{equation}
\end{definition}

\begin{remark}
\label{remark:ConvergenceProofs}
Defining $R'$ to map into $\R$ is only possible if the sum in the first case in~\eqref{eq:abstr-MRM} converges.
We mechanise in Isabelle/HOL that this is indeed the case, which is \Cref{thm:evt-conv}, from which follows that expected rewards until hitting a goal set in terms of their defining sums converge.
Our mechanisation of this convergence fact is, to the best of our knowledge, the first in an interactive theorem prover in terms of lightweight finite-prefix and real analysis semantics.
Convergence of such expected rewards was established in H\"{o}lzl's work~\cite{Hol17}, in the measure-theoretic semantics, using properties of random variables on probability spaces.
This convergence fact is crucial in analysing what happens in the case of multiple path abstractions, most specifically the monotonic absorption property of path abstraction for MRMs.
\end{remark}

\begin{theorem}[\texttt{Abstr.thy/MRM{\_}abstr{\_}well{\_}defined}]
\label{thm:PA-MRM-well-defined}\\
If $\M=(S,s_0,\P,R)$ is an MRM in the free monoid view of \Cref{def:MRM-free-monoid}, then so is $\abstr(A,\M)$, for any $A\subseteq S$.
\end{theorem}

\begin{theorem}[\texttt{Abstr.thy/MRM{\_}monotonic{\_}absorption}]
\label{thm:PA-MRM-monabs}\\
Let $\M=(S,s_0,\P,R)$ be a free-monoid MRM as per \Cref{def:MRM-free-monoid} and~\Cref{conv:substochastic}.
If $A\subseteq B\subseteq S$, we have $\abstr(B,\abstr(A,\M))=\abstr(B,\M)$.
\end{theorem}

\subsection{Path Abstraction as PMC}
\label{sec:PathAbstractionAsPMC}

Now, we argue why a single application of path abstraction allows us to compute reachability probabilities and expected rewards as directly as one could wish, illustrated by a further refinement of the MRM in \Cref{fig:prob-prog-MRM-abstr} as an example.
Then, we briefly show why path abstraction relative to abstraction sets that do not contain goal states preserves reachability probabilities and expected rewards, under the assumption that there is no interaction between the goal states---\ie there are no transitions of positive probability between them, using the later result that path abstraction on MRMs satisfies a monotonic absorption property.

We take an MRM $\M=(S,s_0,\P,R)$ and a set of goal states $G\subseteq S$. We assume that the goal states do not interact; \ie there do not exist states $s,t\in G$ with $s\neq t$ for which $\P(st)>0$.
For example, we can assume that no goal state has outgoing probability, as in our example in \Cref{fig:prob-prog-MRM},
or, as is often done in PMC if the modelled DTMC has to be strictly stochastic, that the goal states are \emph{absorbing}: $\P(ss)=1$ for all $s\in G$. In both cases, no goal state has outgoing probability to a non-goal state.
We now examine what happens if we abstract $\M$ over the largest set that makes sense, \ie the set of all non-goal states $S-G$.
In other words, we compute $\M'=(S,s_0,\P',R')$ where $(S,s_0,\P',R')=\abstr(S-G,\M)$.
We have $\inp(M,S-G)=\set{s_0}$, so the only transitions of positive probability in $\M'$ are (at most the set) $\set{(s_0,g)\colon g\in G}$, each one having, following~\eqref{eq:char-P'(st)}, probability
$$\P'(s_0g)=\P(s_0(S-G)^*g)=\sum\nolimits_{x\in s_0(S-G)^*g}\P(x),$$
where the last equality follows from \Cref{def:prob-lang} as
$$(s_0(S-G)^*g)_\perp=s_0(S-G)^*g.$$
Summing over $g\in G$ then yields
\begin{equation}
\P'(\diamondsuit G)=\sum_{g\in G}\P'(s_0g)=\sum_{x\in s_0(S-G)^*G}\P(x)
=\P(\diamondsuit G),
\end{equation}
where the last equality follows from~\eqref{eq:reach-prob-exp-rew-infsums}. In other words, the probabilities of the transitions in the abstracted MRM $\M'=\abstr(S-G,\M)$ sum up to the exact reachability probability $\P(\diamondsuit G)$ in the original MRM.
The difference from $\M$ is that the total reachability probability to $G$ is now encoded in the most elementary way possible.

Since $\M'$ is an easy-to-understand MRM, analogously to calculating the total reachability probability to reach $G$ there, we can calculate the total expected reward by multiplying probability by reward for each transition, and then summing those over $g\in G$.
Given $g\in G$, we have
\begin{equation}
\P'(s_0g)R'(s_0g)=\sum\nolimits_{x\in s_0(S-G)^*g}\P(x)R(x)
\end{equation}
by~\eqref{eq:abstr-MRM} and the characterisation of the inverse image of $\red(A)$ on two-state sequences in~\eqref{eq:red-char}. It follows that
\begin{equation}
\begin{split}
\E'(\diamondsuit G)&=\sum\nolimits_{g\in G}\P'(s_0g)R'(s_0g)
\\&=\sum\nolimits_{x\in s_0(S-G)^*G}\P(x)R(x)=\E(\diamondsuit G),
\end{split}
\end{equation}
where $\E'(\diamondsuit G)$ denotes the expected reward to reach $G$ in $\M'$, and the final equality in the above follows from~\eqref{eq:reach-prob-exp-rew-infsums}.

We have thus established that abstracting MRMs over the set of non-goal states preserves reachability probabilities and expected rewards.
If we establish \Cref{thm:PA-MRM-well-defined} and \Cref{thm:PA-MRM-monabs}, then we have proven that an abstraction over \emph{any} set of non-goal states preserves these quantities as well.

\begin{figure}
\centering
\begin{tikzpicture}
\draw
(1, 0) node(ghost){}
(2, 0) node[squared node, inner xsep=2pt](s00){$0,0$}
(4, 1) node[squared node, inner xsep=2pt](s22){$2,2$}
(4, -1) node[squared node, inner xsep=2pt](s2inf){$2,\infty$};
\draw
(3.1, 1.3) node{$\frac{5}{9};\frac{157}{15}$}
(3.15, -0.35) node{$\frac{4}{9};\frac{35}{3}$};
\draw[-stealth] (ghost) -- (s00);
\draw[-stealth] (s00) to[bend left = 20] (s22);
\draw[-stealth] (s00) to[bend left = -20] (s2inf);
\end{tikzpicture}
\caption{The MRM from \Cref{fig:prob-prog-MRM-abstr} fully model checked}
\label{fig:prob-prog-MRM-abstr-fully}
\end{figure}
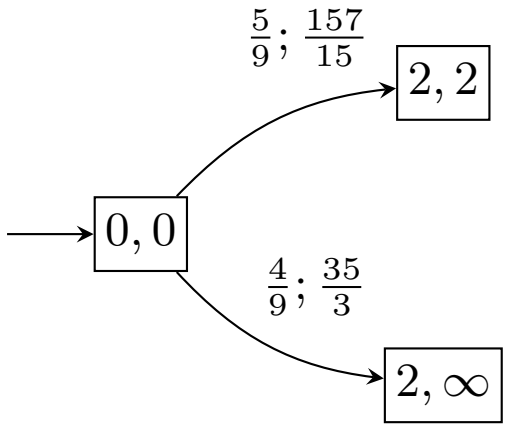

\begin{example}
One can further abstract the MRM in \Cref{fig:prob-prog-MRM-abstr} over the set of non-terminal states $S-\set{(2,2),(2,\infty)}$ to obtain the MRM in \Cref{fig:prob-prog-MRM-abstr-fully}, where we do not draw states that become isolated.
We can read off that the probability $\P(\diamondsuit\set{(2,2),(2,\infty)})$ of termination is $\frac{5}{9}+\frac{4}{9}=1$, thus the program terminates almost surely.
Furthermore, the \emph{conditional} expected time to reach the state corresponding to the division by zero \emph{given} that we will end up in this state can be directly read off as $\E(\diamondsuit(2,\infty))=\frac{35}{3}$; the total expected time to reach termination is calculated as $\frac{5}{9}\cdot\frac{157}{15}+\frac{4}{9}\cdot\frac{35}{3}=11$, which is indeed higher than the previously established bound in~\eqref{eq:exp-val-lower-bnd}.
Then, monotonic absorption as established in \Cref{thm:PA-MRM-monabs} enables us to conclude that the fully model checked MRM in \Cref{fig:prob-prog-MRM-abstr-fully} could also have been obtained by abstracting the initial MRM in \Cref{fig:prob-prog-MRM} over the set of non-terminal states directly.
\end{example}

\section{Verified Path Abstraction}
\label{sec:verified-path-abstraction}

Our formalisation work is as follows.
We
\begin{list}{}{\settowidth{\labelwidth}{(\ref{sec:MechanisingMonotonicAbsorption})}\setlength{\leftmargin}{\labelwidth}\addtolength{\leftmargin}{\labelsep}}
\item[(\ref{sec:MRMsInIsabelle})] formalise \emph{DTMCs} and \emph{MRMs} (\Cref{def:MRM-free-monoid},~\Cref{conv:substochastic}) and
\item[(\ref{sec:FormalisingPathAbstraction})] the \emph{path abstraction algorithms} (\Cref{def:abstr,def:abstr-MRM}),
\item[(\ref{sec:MechanisingWellDefinedness})] mechanise the algorithms' \emph{well-definedness} (\Cref{thm:PA-MC-well-defined,thm:PA-MRM-well-defined}), and
\item[(\ref{sec:MechanisingMonotonicAbsorption})] mechanise the \emph{monotonic absorption} property of these algorithms\\ (\Cref{thm:PA-MC-monabs,thm:PA-MRM-monabs}).
\end{list}
We use Isabelle's standard \texttt{HOL-Library.Sublist} to represent and reason with words, languages, and free monoids, and
\texttt{HOL-Analysis.Infinite\_Sum} to reason about infinite sums, which uses \Cref{def:infsum} as the definition of infinite sums.

\newcommand{\problang}{\textit{'a lang $\Rightarrow$ real}}

\subsection{MRMs in Isabelle}
\label{sec:MRMsInIsabelle}

Our formalisation design of \Cref{def:MRM-free-monoid} and~\Cref{conv:substochastic} in Isabelle/HOL is as follows.
We define the type synonyms
\begin{isaalign*}
\textit{'a lang} &= \textit{``'a list set''} \\
\textit{'a dtmc} &= \textit{``'a set $\times$ 'a $\times$ }(\textit{'a lang $\Rightarrow$ real})\textit{''}
\end{isaalign*}
and define our class of DTMCs \textit{DTMC :: ``'a dtmc set''}, through
\begin{isaalign*}
	\textbf{\text{de}}&\text{\textbf{finition}}\textit{ DTMC :: ``'a dtmc set'' }\textup{\textbf{where}}\\
	&\textit{``DTMC $\equiv$ }\{(\textit{S, s0, P})\textit{. finite S $\land$ s0 $\in$ S $\land$ P $\in$ problang}\}\textit{''}
\end{isaalign*}
where \textit{problang :: 'a lang $\Rightarrow$ real} represents transition probability functions independent of state space through singleton languages, embedding the natural extension to languages of~\eqref{eq:prob-lang}.
Its definition involves several auxiliary functions; for clarity, we therefore prove a characterisation lemma for the class \textit{problang}:
\begin{isaalign}
\text{\textbf{le}}&\text{\textbf{mma} }\textit{problang\_char: ``problang $=$ }\textup{\{}\textit{P. }(\textit{$\exists$A. finite A}\label{problang-char}\\
&\land(\textit{$\forall$xs. }(\textit{xs $\notin$ lists A $\land$ length xs $\geqslant2$ $\longrightarrow$ P }\textup{\{}\textit{xs}\textup{\}}\textit{ $=0$})))\notag\\
&\land\textit{P }\{[]\} = 0 \land (\forall\textit{s}.\,\textit{P }\{[\textit{s}]\} = 1) \land (\forall\textit{xs}.\,\textit{P }\{\textit{xs}\} \geqslant 0)\notag\\
&\land(\forall\textit{s}.\, \textit{sum }(\lambda\textit{t}.\,\textit{P }\{[\textit{s}, \textit{t}]\})\,\{\textit{t}.\,\textit{P }\{[\textit{s}, \textit{t}]\} \neq 0\} \leqslant 1)\notag\\
&\land(\forall \textit{ xs s ys. P }\{\textit{xs }@\textit{ s }\#\textit{ ys}\} = \textit{P }\{\textit{xs }@\,[\textit{s}]\} * \textit{P }\{\textit{s }\#\textit{ ys}\})\notag\\
&\land(\forall\textit{L. P L }=\textit{ infsum }(\lambda\textit{xs. P }\{\textit{xs}\})\,(\textit{lminpfx L}))\}\textit{''}\notag
\end{isaalign}
Here, \textit{lminpfx :: ``'a lang $\Rightarrow$ 'a lang''} is Isabelle's counterpart of taking the prefix-minimal part $L_\perp$ of a language $L$:
\begin{isaalign*}
\text{\textbf{de}}&\text{\textbf{finition} }\textit{lminpfx :: \!``'a lang $\Rightarrow$ 'a lang'' }\text{\,\textbf{where}\,}\textit{``lminpfx L}\\
&\equiv \{\textit{xs }\in\textit{ L. }\forall\textit{xs' }\in\textit{ L. prefix xs' xs }\longrightarrow\textit{ xs' }=\textit{ xs}\}\textit{''}
\end{isaalign*}

\begin{remark}
The most natural way to encode a probability function on the free monoid of a state space is to assign zero probability to paths of length $\geqslant2$ that contain states outside of the state space, to assign probability $1$ to \emph{any} path of length $1$, and to assign probability $0$ to the empty word.
The problem with encoding an explicit state space is that each time we need to chain the condition that some probability function is inside a class for access to to-be-proven convergence lemmas, we need to supply the state space.
To prevent these parametric conditions, we only encode that a finite state space confining the probability function to zero-ness on paths of length $\geqslant2$ that do not strictly run within the state space \emph{exists}, cf.\ the first clause in the set comprehension of~\eqref{problang-char}.
This saves considerable boilerplate.
The main results still apply in case one wants to restrict the probability function \textit{P} in a triple (\textit{S, s0, P}) to be zero on lists of length $\geqslant2$ that are not in \textit{lists S}, as adding more conditions to a lemma breaks neither its proof nor its conclusion.
\end{remark}

We now define \textit{MRM :: ``$($'a dtmc $\times$ $($'a list $\Rightarrow$ real$))$ set''} as the class of MRMs, thus incorporating a reward function that acts on lists directly:
\begin{isaalign}
\text{\textbf{de}}&\text{\textbf{finition} }\textit{MRM :: ``$($'a dtmc $\times$ $($'a list $\Rightarrow$ real$))$ set''}\label{MRM}\\
&\text{\textbf{where} }\textit{``MRM $=$ }\textup{\{((}\textit{S, s0, P}\textup{)}\textit{, R}\textup{)}\textit{. }\textup{(}\textit{S, s0, P}\textup{)}\textit{ $\in$ DTMC}\notag\\
&\land\textup{ (}\forall\textit{xs. length xs $\leqslant 1$ $\lor$ P }\textup{\{}\textit{xs}\textup{\}}\textit{ $=0$ $\longrightarrow$ R xs $=0$}\textup{)}\notag\\
&\land\textup{ (}\forall\textit{xs a ys. P }\textup{\{}\textit{xs }@\textit{ a }\#\textit{ ys}\textup{\}}\textit{ $\neq0$ $\longrightarrow$}\notag\\
&\textit{R }\textup{(}\textit{xs }@\textit{ a }\#\textit{ ys}\textup{)}\textit{ = R }(\textit{xs }@\,[\textit{a}]\}) + \textit{R }(\textit{a }\#\textit{ ys}))\textup{\}}\textit{''}\notag
\end{isaalign}
and thus directly specifying a reward function \textit{R} on paths, following \Cref{def:MRM-free-monoid}.

\subsection{Formalising the Path Abstraction Algorithm}
\label{sec:FormalisingPathAbstraction}

We have specified two path abstraction algorithms in \Cref{def:abstr} and \Cref{def:abstr-MRM}, respectively, where the former can be immediately formulated in terms of the definition of MRMs (\Cref{def:MRM-free-monoid}), and the latter uses the former.
We first formalise the algorithm for DTMCs:
\begin{isaalign}
\text{\textbf{de}}&\text{\textbf{finition}}\textit{ abstr :: ``'a set $\Rightarrow$ 'a dtmc $\Rightarrow$ 'a dtmc''}\label{isabelle-abstr}\\
&\text{\textbf{where} }\textit{``abstr A M $\equiv$ }(\textit{fst M, fst }(\textit{snd M})\textit{,}\notag\\
&\lambda\textit{L. }(\textit{$\textstyle\sum\nolimits_\infty$ xs $\in$ lminpfx L. }(\text{\textbf{if}}\textit{ xs $=$ $[]$ }\text{\textbf{then} }0\text{\textbf{ else}}\notag\\
&(\text{\textbf{if} }\textit{length xs $=1$ }\text{\textbf{then} }1\text{ \textbf{else}}\notag\\
&(\text{\textbf{if} }(\textit{xs $\in$ lists }((\textit{fst M}) - (\textit{A }- \textit{inp M A})))\notag\\
&\text{\textbf{then} }(\textit{snd }(\textit{snd }(\textit{M}))\textit{ }(\textit{vimage }(\textit{red A})\textup{ \{}\textit{xs}\textup{\}})\text{\textbf{ else }}0)))))\notag
\end{isaalign}
We now specify \textit{inp} and \textit{red}, Isabelle's counterparts of \Cref{def:inp} and \Cref{def:red}. For \textit{inp}, we have:
\begin{isaalign*}
	\text{\textbf{de}}&\text{\textbf{finition }}\textit{inp :: 		``'a dtmc}\Rightarrow\textit{'a set}\Rightarrow\textit{		'a set'' }\text{\textbf{where}}\\
	&\textit{``inp M A }\equiv\{\textit{t }\in\textit{ A. t }= (\textit{fst }(\textit{snd M}))\\
	&\lor(\textit{snd }(\textit{snd M}))\,\{[\textit{s}, \textit{t}] \mid \textit{ s}.
	\textit{ s }\in(\textit{fst M}) - \textit{A}\}
	\neq 0\}\textit{''}
\end{isaalign*}
The function \textit{red} is defined through an auxiliary function, so for understandability we specify it with a characterisation:
\begin{isaalign*}
\text{\textbf{le}}&\text{\textbf{mma} }\textit{red\_char:}\\
&\textit{``}\textit{red A }[] = []\textit{'' ``red A }[\textit{x}] = [\textit{x}]\textit{''}\\
&\textit{``red A }(\textit{x } \# \textit{ y } \# \textit{ xs}) = (\text{\textbf{if} }(\textit{x} \in \textit{A} \land \textit{y} \in \textit{A})\\
&\text{\textbf{then} }\textit{red A }(\textit{x }\#\textit{ xs}) \text{ \textbf{else} }\textit{x }\#\textit{ red A }(\textit{y }\#\textit{ xs}))\textit{''}
\end{isaalign*}
We are now ready to formalise the path abstraction algorithm for MRMs, as follows:
\begin{isaalign}
\text{\textbf{de}}&\text{\textbf{finition} }
\textit{MRM\_abstr :: ``'a set $\Rightarrow$ }(\textit{'a dtmc $\times$ }(\textit{'a list}\label{isabelle:MRM-abstr}\\
&\textit{$\Rightarrow$ real}))\textit{ $\Rightarrow$ }(\textit{'a dtmc $\times$ }(\textit{'a list $\Rightarrow$ real}))\textit{''}\text{ \textbf{where}}\notag\\
&\textit{``MRM\_abstr A MR $\equiv$}\text{ \textbf{case} }\textit{MR}\text{ \textbf{of} }(\textit{M, R})\textit{ $\Rightarrow$}\notag\\
&(\text{\textbf{let} }(\textit{S, s0, P})\textit{ $=$ M; }(\textit{S, s0, Q})\textit{ $=$ abstr A M }\text{\textbf{in}}\notag\\
&(\textit{abstr A M, $\lambda$xs. }(\text{\textbf{if} }(\textit{length xs $> 1\land$ Q }\textup{\{}\textit{xs}\textup{\}}\textit{ $\neq0$})\notag\\
&\text{\textbf{then} }((\textit{1 / Q }\textup{\{}\textit{xs}\textup{\}})\textit{ * infsum }(\textit{$\lambda$ys. P }\textup{\{}\textit{ys}\textup{\}}\textit{ * R ys})\notag\\
&(\textit{lminpfx }(\textit{vimage }(\textit{red A})\textup{ \{}\textit{xs}\textup{\}})))\text{ \textbf{else} }0)))\textit{''}\notag
\end{isaalign}
Then, the underlying DTMC of an MRM gets mapped to its abstraction after applying \textit{abstr}; the new reward function is specified by mapping a path to the (conditional) expected reward on the abstract path defined by \textit{vimage }(\textit{red A}) \{\textit{xs}\}, conforming to \Cref{def:MRM-free-monoid}.

\subsection{Mechanising Well-Definedness}
\label{sec:MechanisingWellDefinedness}

We now describe the mechanisation of the well-definedness properties of the path abstraction algorithms for DTMCs and MRMs, respectively, \ie \Cref{thm:PA-MC-well-defined,thm:PA-MRM-well-defined}.
Their Isabelle counterparts are, respectively,~\eqref{isabelle-abstr-well-defined} and~\eqref{isabelle-MRM-abstr-well-defined}:
\begin{isaalign}
\text{\textbf{le}}&\text{\textbf{mma}}\textit{ abstr\_well\_defined:}\label{isabelle-abstr-well-defined}\\
&\text{\textbf{assumes} }\textit{``M $\in$ DTMC'' }\textit{``A $\subseteq$ fst M''}\notag\\
&\text{\textbf{shows} }\textit{``abstr A M $\in$ DTMC''}\notag
\end{isaalign}
\begin{isaalign}
\text{\textbf{le}}&\text{\textbf{mma} }\textit{MRM\_abstr\_well\_defined:}\label{isabelle-MRM-abstr-well-defined}\\
&\text{\textbf{assumes }}\textit{``}(\textit{M}, \textit{R}) \in \textit{MRM'' ``M }= (\textit{S}, \textit{s0}, \textit{P})\textit{''}\textit{ ``A} \subseteq \textit{S''}\notag\\
&\text{\textbf{shows} }\textit{``MRM\_abstr A }(\textit{M}, \textit{R}) \in \textit{MRM''}\notag
\end{isaalign}
We now give the proof outlines of those, in the notation of their mathematical counterparts \Cref{thm:PA-MC-well-defined,thm:PA-MRM-well-defined}. In these outlines, we indicate where proof engineering complexity in Isabelle diverges from conceptual complexity to highlight the mechanisation contributions.

\begin{proof}[Proof outline of \Cref{thm:PA-MC-well-defined}]
\label{proof-outline-PA-MC-well-defined}
	We follow the pen-and-paper proof in~\cite[App.~B]{HM25x}.
	Thus, let $M=(S,s_0,\P)$ be a DTMC, where $S$ is a 		finite set, $s_0\in S$, and $\P:\U^*\to\R$ is such that $\P$
	is in the right-hand set of~\eqref{problang-char}, where we fix $\U$ as the universal set of symbols.
	For simplicity, we omit the brackets for singleton
	languages.
	Spelled out, (1) there exists a finite set $A$ such that $\P(x)=0$ for all $x\in\U^*-A^*$ with $|x|\geqslant2$; (2) $\P(\varepsilon)=0$; (3) $\P(s)=1$ for all $s\in\U$; (4) $\P:\U^*\to[0,\infty)$; (5) for all $s\in\U$ we have $\sum_{t\in\U\colon\P(st)\neq0}\P(st)\leqslant1$; (6) for all $x,y\in\U^*$ and $s\in\U$ we have $\P(xsy)=\P(xs)\P(sy)$; and (7) for all $L\subseteq\U^*$ we have $\P(L)=\sum_{x\in L_\perp}\P(x)$.
	Write $M'=(S,s_0,\P')$ for the abstracted DTMC $M'=\abstr(A,M)$, where $A\subseteq S$, with $A$ not to be confused with the $A$ from condition (1).
	We are thus done if we can establish that $\P'$ satisfies conditions (1), \ldots, (7) as well.
	Condition (1) is satisfied by $\P'$ if we choose $S$ as our finite set, and then follows directly from~\eqref{isabelle-abstr}.
	Conditions (2) and (3) for $\P'$ also follow immediately from~\eqref{isabelle-abstr}.
	Condition (4) follows directly from (4) for $\P$ and~\eqref{isabelle-abstr} combined.
	Condition (7) for $\P'$ follows immediately from how~\eqref{isabelle-abstr} specifies how $\P'$ behaves on languages.
	At this point, to prove that $\P'$ satisfies conditions (5) and (6), conceptual complexity diverges from proof
	engineering complexity.
	Condition (5) is established for $\P'$ as follows, in \texttt{Abstr.thy/abstr\_well\_defined}.
	Since $\P'$ satisfies (1), $\sum_{t\in\U\colon\P'(st)\neq0}\P'(st)$ is a sum of finitely many numbers.
	For each such $t$ we have $\P'(st)\leqslant\P(\red(A)^{-1}(st))$ by~\eqref{isabelle-abstr}, or, equivalently, by \Cref{def:abstr}.
	But the sum of the $\P(\red(A)^{-1}(st))$'s over these values of $t$ is $\P(L)$ where $L=\biguplus_{t\in\U\colon\P'(st)\neq0}(\red(A)^{-1}(st))_\perp$,
	as $L=L_\perp$.
	We have $L\subseteq sS^+$ as its components are contained in $sS^+$ as well, and we may apply \Cref{lem:prob-lang-finconf-startingpt-conv-to-0-1} to establish that $\P(L)\leqslant1$ and hence $\P'$ satisfies condition (5).
	Condition (6) is then tackled as follows: if $x,y\in\U^+$ and $s\in\U$ are such that $xsy\in(S-(A-\inp(M,A)))^*$, we have $\P'(z)=\P(L_z)$ for each $z\in\set{xs,sy,xsy}$ where $L_z=(\red(A)^{-1}(z))_\perp$.\footnote{Note that $\P(L)=\P(L_\perp)$ for every language $L$, as $\perp$ is idempotent.}
	Then, proving $\P(L_{xsy})=\P(L_{xs})\cdot\P(L_{sy})$ comes down to proving
	\begin{equation}
	\sum\nolimits_{w\in L_{xsy}}\P(w)=\sum\nolimits_{u\in L_{xs}}\P(u)\sum\nolimits_{v\in L_{sy}}\P(v)
	\end{equation}
	which holds because $L_{xsy}=\biguplus_{u\in L_{xs}}u\star L_{sy}$, most of whose foundation we mechanise in \texttt{Freemonoid.thy/\allowbreak ltappend\_\allowbreak lminpfx\_\allowbreak vimage\_\allowbreak red}.  Indeed: $\P$ converges on $L_{xsy}$ because $L_{xsy}$ is contained in a monoid on a finite set and all its words start with the first symbol of $xsy$.
	We may thus use our variant of Tonelli's theorem for non-negative real-valued functions---\Cref{lem:tonelli}---applied to $B=L_{xsy}$ and $I=L_{xs}$---and the convergence result \Cref{lem:prob-lang-finconf-startingpt-conv-to-0-1} for the languages $L_{xs}$, $L_{sy}$, and $L_{xsy}$, which is applicable because the three languages start with the first symbol of $xs$, $sy$, and $xsy$, respectively (note that these words are non-empty by assumption) and are contained in $S^+$.
	The edge cases $x\neq\varepsilon\land y\neq\varepsilon\land xsy\notin (S-(A-\inp(M,A)))^*$ and $x=\varepsilon\lor y=\varepsilon$ are conceptually trivial and also do not pose further mechanisation challenges.
\end{proof}

\begin{theorem}[Convergence Lemma I: \texttt{Problang.thy/\allowbreak has\_\allowbreak csp\_\allowbreak finite\_\allowbreak confinable\_\allowbreak substoch}]
\label{lem:prob-lang-finconf-startingpt-conv-to-0-1}
Given $L\subseteq\U^+$ such that there exist
$a\in\U$ and $L'\subseteq\U^*$ and a finite $A\subseteq\U$ such that $L=aL'$ and $L\subseteq A^+$, a
function $\P$ satisfying $(1), \ldots, (7)$ in the
previous proof outline has the property that $
\P(L)=\sum_{x\in L_\perp}\P(x)$ converges to a real
number in $[0,1]$.
\end{theorem}

\begin{proof}
Conceptually, this follows by monotonicity of convergence of non-negative functions, combined with the observation that $\P(L)\leqslant1$ in case $L$ is finite, which follows by induction on the depth of $L_\perp$, \ie the maximum length $\max_{x\in L_\perp}|x|$ of a word in $L_\perp$.
In Isabelle, one needs to work out these connections explicitly.
We first establish, in
\texttt{Problang.thy/\allowbreak is\_\allowbreak ltree\_\allowbreak substoch},
that $\P(L)\leqslant1$ with the additional assumption that $L$ is finite,
by induction on the depth of $L_\perp$. The key idea is that chopping off the final symbols of the words in $L_\perp$ of maximum length yields $K_\perp$ of another language with depth one less than that of $L_\perp$; by grouping paths we thus establish $\P(L)\leqslant\P(K)$, and we can reach all the way down to the base case $\P(\set{a})=1$.
Then, to prove that $\P(L)$ converges to a real number in $[0,1]$, we use Isabelle's \texttt{nonneg\_\allowbreak bdd\_\allowbreak above\_\allowbreak summable\_\allowbreak on} lemma to prove that $\P(L)$ converges at all, taking as preconditions that $\P$ is a non-negative valued function and the set of its partial sums (\wrt finite subsets) is bounded from above by~$1$.
Then, we use Isabelle's \texttt{infsum\_\allowbreak nonneg\_\allowbreak is\_\allowbreak SUPREMUM\_\allowbreak real} lemma that states that a non-negative function converging on a set converges to the supremum of its partial sums; proving that the latter is in $[0, 1]$ is easy.
\end{proof}

\begin{theorem}[Tonelli's Theorem for non-negative functions: \texttt{Prob\allowbreak lang\allowbreak .thy/\allowbreak double\_\allowbreak sum}]
\label{lem:tonelli}
Let $f:B\to[0,\infty)$ be a function from a set $B$ to the non-negative reals, such that $f$ converges on $B$ to a value $y\in[0,\infty)$: $\sum_{x\in B}f(x)=y$.
Suppose that $B=\biguplus_{i\in I}B_i$ is a partition of $B$, where $I$ is an index set (possibly infinite).
Then,
\begin{equation}
\label{eq:tonelli}
i\mapsto\sum_{x\in B_i}f(x)\text{ converges on }I\text{ and }\sum_{i\in I}\sum_{x\in B_i}f(x)=y.
\end{equation}
\end{theorem}

\begin{proof}
This is Tonelli's theorem for non-negative numbers.
HOL-Analysis provides this result for sums over $\Sigma$-types (\texttt{has\_\allowbreak sum\_\allowbreak SigmaD}), but our language decompositions are disjoint unions; rather than reindexing them as $\Sigma$-types, we prove the union version directly from the bounded partial sums of non-negative functions.
We first prove that $\sum_{i\in J}\sum_{x\in B_i}f(x)\leqslant y$ for every finite $J$, which is done via finite set induction on $J$;
then, we use the same machinery from the \texttt{nonneg\_\allowbreak bdd\_\allowbreak above\_\allowbreak summable\_\allowbreak on} and \texttt{infsum\_\allowbreak nonneg\_\allowbreak is\_\allowbreak SUPREMUM\_\allowbreak real} lemmas to establish the first statement of~\eqref{eq:tonelli} and with that its second.
\end{proof}

\begin{proof}[Proof outline of \Cref{thm:PA-MRM-well-defined}]
Since, to the best of our knowledge, \Cref{thm:PA-MRM-well-defined} is new in our semantic framework, we develop its proof and mechanisation from scratch.
Thus, let $\M=(M,R)$ with $M=(S,s_0,\P)$.
We extend the spirit of the proof outline of \Cref{thm:PA-MC-well-defined}.
Write $\M'=\abstr(A,\M)$, $M'=\abstr(A,M)=(S,s_0,\P')$, and $\M'=(M',R')$---which is due to \Cref{def:abstr-MRM}.
Due to \Cref{thm:PA-MC-well-defined} we already know that $M'$ is a DTMC.
We thus need to verify, conforming to~\eqref{MRM}, that (1) $R'(x)=0$ whenever $x\in\U^*$ satisfies $|x|\leqslant1$ or $\P'(x)=0$, and (2) for all $x,y\in\U^*$ and $s\in\U$ for which $\P'(xsy)\neq0$ we have $R'(xsy)=R'(xs)+R'(sy)$.
The fact that condition (1) holds for $R'$ is immediately due to~\eqref{isabelle:MRM-abstr}.
Condition (2) is established to hold for $R'$ as follows.
Write $L_z=(\red(A)^{-1}(z))_\perp$ again, for $z\in\set{xs,sy,xsy}$.
We first consider the case that $x$ and $y$ are non-empty and $\P'(xsy)\neq0$ (which implies $xsy\in(S-(A-\inp(M,A)))^*$).
Then, $\P'(xsy)R'(xsy)=\sum_{w\in L_{xsy}}\P(w)R(w)$, $\P'(xsy)R'(xs)=\sum_{v\in L_{sy}}\P(v)\sum_{u\in L_{xs}}\P(u)R(u)$, and ulti\-mate\-ly
$\P'(xsy)R'(sy)=\sum_{u\in L_{xs}}\P(u)\sum_{v\in L_{sy}}\P(v)R(v)$, because we have $\P'(xsy)=\P'(xs)\P'(sy)$ from \Cref{thm:PA-MC-well-defined}.
We thus wish to prove
\begin{align}
\sum\nolimits_{w\in L_{xsy}}\P(w)R(w)&=\sum\nolimits_{v\in L_{sy}}\P(v)\sum\nolimits_{u\in L_{xs}}\P(u)R(u)\label{eq:double-sums-well-definedness-PA-MRM}
\\&+\sum\nolimits_{u\in L_{xs}}\P(u)\sum\nolimits_{v\in L_{sy}}\P(v)R(v).\notag
\end{align}
Using the property $L_{xsy}=\biguplus_{u\in L_{xs}}u\star L_{sy}$ we rewrite the first sum as $$\sum\nolimits_{u\in L_{xs}}\sum\nolimits_{v\in L_{sy}}\P(u)\P(v)(R(u)+R(v)),$$ which is possible as $L_{xs}\subseteq \U^*s$ and $L_{sy}\subseteq s\U^*$, which we also mechanise, so that we may apply $\P(u\star v)=\P(u)\P(v)$ and $R(u\star v)=R(u)+R(v)$ because $\M=(S,s_0,\P,R)$ is an MRM.
Redistributing terms gives~\eqref{eq:double-sums-well-definedness-PA-MRM}.
However, redistributing sums in interaction with Isabelle's HOL-Analysis library is only allowed if we can prove that all summands converge, and all of a sudden \Cref{lem:prob-lang-finconf-startingpt-conv-to-0-1} no longer suffices and we need to invent new machinery.
The trick is to prove that, for a language $L$ of the form $L_z=(\red(A)^{-1}(z))_\perp$ for $z\in\U^*$, the sum $\sum_{w\in L_z}\P(w)R(w)$ converges in case $R$ is of the following special form: $R(z)=|z|_{st}$, where $s,t\in S$, which is the number of times $st$ appears as an edge in the path $z$---cf.\ the formal definition of \emph{factor occurrence} in \Cref{subsec:lang-free-monoid}. Such sums are \emph{expected visiting times} of the transition $st$, whose convergence we prove in \Cref{thm:evt-conv}.
It is not (too) hard to prove that for general $R$ we have the identity $R(z)=\sum_{s,t\in\sett(z)}|z|_{st}\cdot R(st)$, and we have expressed $R(z)$ as a linear combination of the values of $R$ on the two-state factors of $z$, which corresponds to computing a weighted sum of transition rewards \wrt how many times a transition occurs in a path.
Given that, as $w,u,v$ range over their indexing sets in~\eqref{eq:double-sums-well-definedness-PA-MRM}, we have $\sett(w),\sett(u),\sett(v)\subseteq S$, we may thus express each of $R(w)$, $R(u)$, $R(v)$ as a linear combination of the values $R(st)$ for $s,t\in S$.
Then, the convergence result \Cref{thm:evt-conv} can be pushed to transform~\eqref{eq:double-sums-well-definedness-PA-MRM} for the special case $R$ is of the form $R(z)=|z|_{st}$ (where $s,t\in S$) to the general case, and a smooth interaction with the HOL-Analysis library, in combination with the same instance of Tonelli's theorem, \Cref{lem:tonelli}, applied to $B=L_{xsy}$ and $I=L_{xs}$, follows.
The remaining edge cases for $x$ and $y$ are easy, both conceptually and in proof engineering complexity.
\end{proof}

\begin{theorem}[Convergence Lemma II: \texttt{Problang.thy/\allowbreak evt\_\allowbreak summable\_\allowbreak vimage\_\allowbreak red\_\allowbreak all}]
\label{thm:evt-conv}
Let $x\in\U^*$.
Let $\P:\U^*\to\R$ satisfy the defining conditions of~\eqref{problang-char}.
Let $A\subseteq\U$ be a finite set of symbols.
Write $L_x=(\red(A)^{-1}(x))_\perp$. Let $a,b\in\U$.
Then, the following sum converges:
\begin{equation}
\label{eq:evt-conv}
\sum\nolimits_{y\in L_x}|y|_{ab}\cdot\P(y)
\end{equation}
\end{theorem}
\begin{proof}
Since the terms of~\eqref{eq:evt-conv} are non-negative, our goal comes down to giving a proper upper bound on the sum.
Intuitively, as $y$ becomes larger, probabilities of (added) transitions are multiplied while new occurrences of $ab$ are only added, and we either ``leak'' probability constantly when looping through $ab$ or the language is finite or the probabilities of paths are zero beyond visiting $ab$ as a transition more than once.
We proceed by a case distinction on $x$. For $|x|\leqslant1$, $L_x$ is finite and the sum in~\eqref{eq:evt-conv} trivially converges.
For $x=aby$ with $a,b\in\U$ and $y\in\U^*$, we proceed with reverse induction on $y$.
The case $y=\varepsilon$ comes down to proving that the sum of~\eqref{eq:evt-conv} converges for $|x|=2$, which we prove in a separate lemma.
Now, if~\eqref{eq:evt-conv} converges for some
$x=aby$, let us prove that this is the case if we
replace $x$ by $abzc$ where $z\in\U^*$ and
$c\in\U$. Since $|abz|\geqslant2$, we can write
$abz$ as $wd$ where $w\in\U^*$ and $d\in\U$.
Then, we have $x=wdc$ as our new~$x$.
Since $L_{wdc}=L_{wd}\star L_{dc}$ and $L_{wd}\subseteq\U^*d$ and $L_{dc}\subseteq d\U^*$ and~\eqref{eq:evt-conv} converges for $x=wd$ by the induction hypothesis, we may use \Cref{thm:evt-conv-path-conc} applied to $K=L_{wd}$ and $L=L_{dc}$ and the fact that~\eqref{eq:evt-conv} converges for the two-letter word $x=dc$.
This is mechanised in \texttt{Problang.thy/\allowbreak evt\_\allowbreak summable\_\allowbreak vimage\_\allowbreak red\_\allowbreak 2}.
The latter fact is seen as follows.
Following the characterisation~\eqref{eq:red-char}, $L_{dc}$ is infinite only if $d\in A\land c\notin A$, in which case $L_{dc}=dA^*c$.
If $a\notin A\lor b\notin A$, then $|y|_{ab}\leqslant1$ for all $y\in L_{dc}$ and we have convergence.
The interesting case is thus $a\in A\land b\in A$.
Defining $U=\set{v\in A\colon \P(vA^*c)>0}$,
which intuitively is the set of all states in $A$ such that $c$ is reachable from a path starting in $v$ moving through $A$, with positive probability, we look at two cases.
If $a\notin U\lor b\notin U$, then one can show that for all $y\in L_{dc}$ with $|y|_{ab}>0$ we have $\P(y)=0$ and we have again convergence.
For $a\in U\land b\in U$,
we have $\P(bA^*c)>0$ by definition of $U$ and thus the set $D=\set{x\in bA^*c\colon\P(x)>0}$ is non-empty, as $(bA^*c)_\perp=bA^*c$, as $c\notin A$.
Let $y_-\in D$ be of minimal length, and define the language
$$
\begin{aligned}
H\equiv\ &\set{y\colon\exists y'.\,y'\in L_{dc}\land y\sqsubseteq y'}\\
&\,\cap((b\U^*\cap\U^*ab)-\U^*ab\U^+).
\end{aligned}
$$
Clearly, $y_-\notin H$ because $b\neq c$. Now, $H\uplus\set{y_-}$ is a language contained in the free monoid on a finite set and has a common starting point $b$, so that we may use \Cref{lem:prob-lang-finconf-startingpt-conv-to-0-1} to conclude that $\P(H\uplus\set{y_-})\leqslant1$.
Since $(H\uplus\set{y_-})_\perp=H\uplus\set{y_-}$ (because $y_-\in D$ is of minimal length),
we can use $\P(y_-)>0$ to deduce that $0\leqslant p<1$ if we set $p\equiv\P(H)$.
We now prove that $\P(\set{y\in L_{dc}\colon |y|_{ab}=n})\leqslant p^{n-1}$ for $n\in\N^+$, which comes with its own set of mechanisation challenges that we do not elaborate on here.
It thus follows that, for $x=dc$, the sum in~\eqref{eq:evt-conv} can be bounded as
\begin{equation}
\sum\nolimits_{y\in L_{dc}}|y|_{ab}\cdot\P(y)
\leqslant \sum\nolimits_{n=1}^\infty n\cdot p^{n-1}
\end{equation}
and this upper bound converges because $0\leqslant p<1$, and does so to $(1-p)^{-2	}$, which is a standard result in undergraduate calculus and real analysis.
\end{proof}

\begin{theorem}[\texttt{Problang.thy/\allowbreak evt\_\allowbreak summable\_\allowbreak path\_\allowbreak concatenation}]
\label{thm:evt-conv-path-conc}
Given $K,L\subseteq\U^*$, $\sum_{x\in K\star L}|x|_{ab}\cdot\P(x)$ converges if $\sum_{x\in K}|x|_{ab}\cdot\P(x)$ and $\sum_{x\in L}|x|_{ab}\cdot\P(x)$ do (to, say, $L_1$ and $L_2$, respectively), whenever $K=K_\perp$ and $L=L_\perp$ and there exist $c,d\in \U$ such that $K\subseteq c\U^*\cap\U^*d$ and $L\subseteq d\U^*$.
\end{theorem}

\begin{proof}
The trick is to prove that for every finite $F\subseteq K\star L$ we have $\sum_{x\in F}|x|_{ab}\cdot\P(x)\leqslant L_1+L_2$;
applying the standard machinery from the \texttt{nonneg\_\allowbreak bdd\_\allowbreak above\_\allowbreak summable\_\allowbreak on} and \texttt{infsum\_\allowbreak nonneg\_\allowbreak is\_\allowbreak SUPREMUM\_\allowbreak real} lemmas again proves that $\sum_{x\in K\star L}|x|_{ab}\cdot\P(x)$ converges.
\end{proof}

\subsection{Mechanising Monotonic Absorption}
\label{sec:MechanisingMonotonicAbsorption}

We prove \Cref{thm:PA-MC-monabs,thm:PA-MRM-monabs}, the monotonic absorption properties.
Their Isabelle counterparts are, respectively:
\begin{isaalign}
\text{\textbf{le}}&\text{\textbf{mma} }\textit{monotonic\_absorption:}\\
&\text{\textbf{assumes} }\textit{``M }\in\textit{ DTMC'' ``A }\subseteq\textit{ B'' ``B }\subseteq\textit{ fst M''}\notag\\
&\text{\textbf{shows} }\textit{``abstr B }(\textit{abstr A M}) = \textit{abstr B M''}\notag
\end{isaalign}
\begin{isaalign}
\text{\textbf{le}}&\text{\textbf{mma} }\textit{MRM\_monotonic\_absorption:}\\
&\text{\textbf{assumes} }\textit{``}(\textit{M}, \textit{R}) \in \textit{MRM}\textit{'' }\textit{`` M }= (\textit{S}, \textit{s0}, \textit{P})\textit{'' } \textit{``A }\subseteq\textit{ B''}\notag\\
&\textit{``B }\subseteq\textit{ S'' }\text{\textbf{shows} }\textit{``MRM\_abstr B }(\textit{MRM\_abstr A }(\textit{M}, \textit{R}))\notag\\
&= \textit{MRM\_abstr B }(\textit{M}, \textit{R})\textit{''}\notag
\end{isaalign}

\begin{proof}[Proof outline of \Cref{thm:PA-MC-monabs}]
We follow the pen-and-paper proof in~\cite[App.~B]{HM25x}.
Thus, let $M=(S,s_0,\P)$ be a DTMC, and let $A\subseteq B\subseteq S$.
Write $M_A=(S,s_0,\P_A)$, $M_B=(S,s_0,\P_B)$, and $M_{AB}=(S,s_0,\P_{AB})$ for the abstractions $M_A=\abstr(A,M)$, $M_B=\abstr(B,M)$, and $M_{AB}=\abstr(B,\abstr(A,M))$, respectively.
To prove $M_{AB}=M_B$, it thus suffices to prove $\P_{AB}=\P_B$.
We fix an $x\in\U^*$ and focus on the case $|x|\geqslant2\land x\in (S-(B-\inp(M,B)))^*$ (the edge cases are easy to verify).
Working out the summation details for $\P_{AB}(x)$ and $\P_B(x)$, repeatedly applying the convergence lemma \Cref{lem:prob-lang-finconf-startingpt-conv-to-0-1}, gives
\begin{align}
\P_{AB}(x)&=\sum\nolimits_{y\in L_x\colon \red(A,y)=y}\sum\nolimits_{z\in K_y}\P(z),\label{eq:PA-MC-monabs1}\\
\P_B(x)&=\sum\nolimits_{z\in L_x}\P(z),\label{eq:PA-MC-monabs2}
\end{align}
where for $w\in\U^*$ we write $K_w\equiv(\red(A)^{-1}(w))_\perp$ and $L_w\equiv(\red(B)^{-1}(w))_\perp$.
We prove
$$L_x=\biguplus\nolimits_{y\in L_x\colon\red(A,y)=y}K_y,$$
whose foundation we mechanise in \texttt{Freemonoid.\allowbreak thy/\\\allowbreak lminpfx\_\allowbreak monab}.
Then, we establish $\P_{AB}(x)=\P_B(x)$ from~\eqref{eq:PA-MC-monabs1} and~\eqref{eq:PA-MC-monabs2}, after applying \Cref{lem:tonelli} again, this time for $B=L_x$ and $I=\set{y\in L_x\colon\red(A,y)=y}$.
\end{proof}

\begin{proof}[Proof outline of \Cref{thm:PA-MRM-monabs}]
Take the notation for $M_A$, $M_B$, and $M_{AB}$ in the previous proof outline---that of \Cref{thm:PA-MC-monabs}, and write $\M=(M,R)$, $\M_A=(M_A,R_A)$, $\M_B=(M_B,R_B)$, and $\M_{AB}=(M_{AB},R_{AB})$ for MRMs such that $\M_A=\abstr(A,\M)$, $\M_B=\abstr(B,\M)$, and $\M_{AB}=\abstr(B,\abstr(A,\M))$.
By \Cref{thm:PA-MC-monabs}, we already have $M_{AB}=M_B$.
We are thus left with establishing that $R_{AB}=R_B$.
Analogously to (the proof of) \Cref{thm:PA-MC-monabs}, working out sums for $\P_{AB}(x)R_{AB}(x)$ and $\P_B(x)R_B(x)$ (for the substantial, non-edge case for $x$) gets us to a to-be-proven equality relating a double sum, coming from the side of $\P_{AB}(x)R_{AB}(x)$, to a single sum, coming from the side of $\P_B(x)R_B(x)$, which again is resolved by the identity $L_x=\biguplus_{y\in L_x\colon\red(A,y)=y}K_y$, which we have formalised in \texttt{Freemonoid.\allowbreak thy/\allowbreak lminpfx\_\allowbreak monab}.
Analogously to (the proof of) \Cref{thm:PA-MRM-well-defined}, we need to express reward functions on paths as linear combinations over transition rewards, and (repeatedly) apply \Cref{thm:evt-conv} for the HOL-Analysis library to allow us to perform the necessary sum manipulations.
\end{proof}

\section{Conclusion}
\label{sec:conclusion}

We have extended model checking and counterexample refinement by path abstraction to MRMs and expected rewards, and lifted it onto a machine-checked foundation.
In Isabelle/HOL, we formalised DTMCs and MRMs in the free monoid view, defined the two path abstraction algorithms, and mechanised their well-definedness and monotonic absorption.
For DTMCs, we mechanised the pen-and-paper proofs of~\cite{HM25x}; for MRMs, both the algorithm and its proofs are new.
Since abstracting over all non-goal states is equivalent to PMC, monotonic absorption implies that path abstraction preserves reachability probabilities and expected rewards for any abstraction set sequence leading to $S - G$.

Our work focuses on the correctness of algorithms rather than on contributing general results to Isabelle's mathematical libraries.
The lightweight finite-prefix semantics of the free monoid view allowed us to work with the infinite sums of HOL-Analysis instead of measure theory.
We expect our convergence lemmas for probability masses and expected visiting times, and our language decomposition lemmas, to be reusable beyond path abstraction.
We hope that the advantages of the free monoid view will lead to its more widespread use in formalisations of Markov models and PMC; and, more generally, that our work motivates carefully choosing the right representation for the ITP problem at hand, even if that representation turns out to be non-standard.

Our choice for Isabelle was partly motivated by our plan to use the Isabelle Refinement Framework~\cite{Lam13,LT12} to refine our formalised algorithms into efficient correct-by-construction implementations for use in a probabilistic model checker.
This was recently done for interval iteration~\cite{KSAHL25}, a standard PMC algorithm that is however monolithic and does not provide explainable counterexamples.
We expect the free monoid view to simplify this process compared to building on the classical view and its existing formalisations:
it is inherently closer to the data structures needed to represent MRMs, and to the linear algebra operations that implement the algorithms' abstract steps.
Beyond this, we plan to mechanise the correspondence between the classical and the free monoid views (\Cref{thm:reach-prob-exp-rew-infsums}), and to extend our approach to further PCTL properties, starting with bounded reachability.

\bibliographystyle{splncs04}
\bibliography{paper.bib}

\clearpage
\appendix
\crefalias{section}{appendix} %
\crefalias{subsection}{appendix} %
\section{Calculation of Quantities in Example MRMs}
We first show, by hand, how one obtains the MRM in~\Cref{fig:prob-prog-MRM-abstr} from the MRM in~\Cref{fig:prob-prog-MRM} when abstracting over the set $A=\set{(1,0),(1,1)}$.
Let us call the respective MRMs $\M$ and $\M'$.
Both states in $A$ are reachable from the outside in one step, thus $\inp(M,A)=A$.
It follows that, in obtaining $\M'$ from $\M$, only the transitions between states in $A$ are removed, and new transitions from $\inp(M,A)$ to the set of states outside of $A$ reachable from $A$ in one step---namely, $\set{(2,0),(2,1)}$---are introduced, which, according to the characterisation in~\eqref{eq:char-P'(st)}, get probabilities equal to the probabilities of the sets
\begin{equation}
\begin{cases}
(1,0)((1,1)(1,0))^*(2,0),&\text{for }((1,0),(2,0)),\\
((1,0)(1,1))^+(2,1),&\text{for }((1,0),(2,1)),\\
((1,1)(1,0))^+(2,0),&\text{for }((1,1),(2,0)),\\
(1,1)((1,0)(1,1))^*(2,1),&\text{for }((1,1),(2,1)).
\end{cases}
\end{equation}
with respect to the transition probability function of the old MRM $\M$, which evaluate to
\begin{equation}
\begin{cases}
(1+\frac{1}{4}+\ldots)\cdot\frac{1}{2}=\frac{2}{3},&\text{for }((1,0),(2,0)),\\
\frac{1}{2}\cdot(1+\frac{1}{4}+\ldots)\cdot\frac{1}{2}=\frac{1}{3},&\text{for }((1,0),(2,1)),\\
\frac{1}{2}\cdot(1+\frac{1}{4}+\ldots)\cdot\frac{1}{2}=\frac{1}{3},&\text{for }((1,1),(2,0)),\\
(1+\frac{1}{4}+\ldots)\cdot\frac{1}{2}=\frac{2}{3},&\text{for }((1,1),(2,1)),
\end{cases}
\end{equation}
where we have used the well-known identity for the \emph{geometric series} $\sum_{n=0}^\infty r^n=\frac{1}{1-r}$ for $-1<r<1$, applied to $r=\frac{1}{4}$.

Now, for the \emph{rewards} for those newly introduced transitions, let us consider the path set $(1,0)((1,1)(1,0))^*(2,0)$ in $\M$ for the transition $((1,0),(2,0))$ in MRM $\M'$.
The paths $(1,0)(2,0)$, $(1,0)(1,1)(1,0)(2,0)$, etc.\ have probabilities and rewards
\begin{equation}
\frac{1}{2};1\quad\frac{1}{8};5\quad\frac{1}{32};9\quad\cdots
\end{equation}
so that the reward for this transition in $\M'$ times its probability $\frac{2}{3}$ becomes
\begin{align*}
&\frac{1}{2}\cdot1+\frac{1}{8}\cdot5+\frac{1}{32}\cdot9+\cdots\\
=&\frac{1}{2}\cdot\left(\left(\sum_{n=0}^\infty\left(\frac{1}{4}\right)^n\right)+4\cdot\sum_{n=0}^\infty n\cdot\left(\frac{1}{4}\right)^n\right)
\\=&\frac{1}{2}\cdot\left(\frac{4}{3}+\frac{16}{9}\right)
\\=&\frac{14}{9}
\end{align*}
so that its reward becomes $\frac{14}{9}/\frac{2}{3}=\frac{7}{3}$,
where we have used the standard identity $\sum_{n=0}^\infty n\cdot r^n=\frac{r}{(1-r)^2}$ for $-1<r<1$, applied to $r=\frac{1}{4}$.
The rewards of the other transitions can be calculated analogously.

To obtain the MRM of~\Cref{fig:prob-prog-MRM-abstr-fully} from that of~\Cref{fig:prob-prog-MRM-abstr}, we first abstract over $\set{(0,0),(0,1)}$---which is similar to the previous abstraction, exploiting the closed forms for $\sum_{n=0}^\infty r^n$ and $\sum_{n=1}^\infty n\cdot r^n$---and then over the full set of non-terminal states $S-\set{(2,2),(2,\infty)}$.
The fact that this gives the fully abstracted or model checked MRM---\ie that of abstracting the MRM in~\Cref{fig:prob-prog-MRM} immediately over its set of non-terminal states---follows from applying the monotonic absorption property (\Cref{thm:PA-MRM-monabs}) twice, for the sequence of state sets
\begin{align*}
A_1&=\set{(1,0),(1,1)},\\ A_2&=\set{(0,0),(0,1)}, \\A_3&=\set{(0,0),(0,1),(1,0),(1,1),(2,0),(2,1)}.
\end{align*}
Indeed: since $A_2\subseteq A_3$ we have
\begin{align*}
&\abstr(A_3,\abstr(A_2,\abstr(A_1,\M)))\\
&\qquad\qquad= \abstr(A_3,\abstr(A_1,\M)).
\end{align*}
Since $A_1\subseteq A_3$ as well, the latter is equal to $\abstr(A_3,\M)$, and the ``biggest'' set $A_3$ absorbs all other state sets.

\end{document}